\documentclass[12pt]{article}
\usepackage{preamble}
\begin{document}
\begin{titlepage}
    \title{Counterfactual and Synthetic Control Method: Causal Inference with Instrumented Principal Component Analysis\thanks{We thank Matteo Lacopini, Emanuele Bacchiocchi for helpful discussion on this paper. There is a Python Package for implementing this method, and a Github repository for a detailed explaination for each key sections, available at \href{https://github.com/CongWang141/JMP.git}{https://github.com/CongWang141/JMP.git}. The repository also contains the latest version, the demonstration code, and the data for the empirical study.}} 

    \author{Cong Wang\thanks{Department of Economics and Law, Sapienza University of Rome.}}
    \date{\today}
    \maketitle
    \begin{center}
        \href{https://github.com/CongWang141/JMP/blob/main/latex/main.pdf}{Job Market Paper, latest version available here.}
    \end{center}

    \begin{abstract}
        \noindent In this paper, we propose a novel method for causal inference within the framework of counterfactual and synthetic control. Matching forward the generalized synthetic control method developed by \cite{xu2017generalized}, our instrumented principal component analysis method instruments factor loadings with predictive covariates rather than including them as regressors. These instrumented factor loadings exhibit time-varying dynamics, offering a better economic interpretation. Covariates are instrumented through a transformation matrix, $\Gamma$, when we have a large number of covariates it can be easily reduced in accordance with a small number of latent factors helping us to effectively handle high-dimensional datasets and making the model parsimonious. Moreover, the novel way of handling covariates is less exposed to model misspecification and achieved better prediction accuracy. Our simulations show that this method is less biased in the presence of unobserved covariates compared to other mainstream approaches. In the empirical application, we use the proposed method to evaluate the effect of Brexit on foreign direct investment to the UK.\\

        \bigskip
    \end{abstract}

    \setcounter{page}{0}
    \thispagestyle{empty}
\end{titlepage}

\pagebreak \newpage
\doublespacing

\section{Introduction} 
\label{sec: introduction}
In this paper, we introduce a novel counterfactual imputation method for causal inference, called the Counterfactual and Synthetic Control method with Instrumented Principal Component Analysis (CSC-IPCA). This method combines the dimension reduction capabilities of Principal Component Analysis (PCA) described by \cite{jollife2016principal} to handle high-dimensional datasets with the versatility of the factor models studied by \cite{bai2003computation}, \cite{bai2009panel}, among others, which accommodate a wide range of data-generating processes (DGPs). The CSC-IPCA method represents a significant advancement over the Generalized Synthetic Control (GSC) method proposed by \cite{xu2017generalized}, which utilizes the Interactive Fixed Effects (IFE) approach to model DGPs and impute missing counterfactuals for causal inference.

The main difference between our method and CSC-IFE\footnote{In this paper, we consider the Generalized Synthetic Control (GSC) method as part of the broader counterfactual and synthetic control framework. Therefore, throughout the paper, we refer to the GSC method as the Counterfactual and Synthetic Control method with Interactive Fixed Effects (CSC-IFE).} lies in how we handle covariates. CSC-IFE combines the structural component $\Lambda_i F_t$ with the regressors $X_{it} \beta$, as shown in the following equation:

\begin{equation}
\label{eqn: ife}
Y_{it} = \Lambda_i F_t + X_{it} \beta + \epsilon_{it}
\end{equation}
Instead of including the covariates $X_{it}$ linearly as regressors, the CSC-IPCA method instruments the factor loadings $\Lambda_{it}$ with predictive covariates through a transformation matrix $\Gamma$. This method is constructed as fallowing: first, it assumes a simple factor model, as in \cite{bai2003computation}, with only the structural component combined with factor loadings $\Lambda_i$ and common factors $F_t$:

\begin{equation}
    \label{eqn: fe}
    Y_{it} = \Lambda_i F_t + \epsilon_{it}
    \end{equation}
    Next, it instruments the static factor loadings $\Lambda_i$ with covariates $X_{it}$, allowing the factor loadings to incorporate time-varying properties and become dynamic:
    
    \begin{equation}
    \label{eqn: instrument}
    \Lambda_{it} = X_{it}\Gamma + H_{it}
\end{equation}

The static factor loadings $\Lambda_i$ in Equation \ref{eqn: fe} are assumed to be time-invariant by most studies in the related literature. By instrumenting the factor loadings $\Lambda_i$ with covariates $X_{it}$ through Equation \ref{eqn: instrument}, we can capture the time-varying properties of the factor loadings. The matrix $\Gamma$, serving as an $L \times K$ mapping function from covariates (with the number of L) to factor loadings (with the number of K), also acts as dimension reduction operation (with $K \ll L$), which aggregates all the information from the covariates into a smaller number of factor loadings, making the model parsimonious. Moreover, this approach represents a novel and effective way for extracting predictive information from covariates within a factor model framework.

The CSC-IPCA method offers several key benefits. First, it inherits the dimension reduction capabilities of conventional PCA method, where the transformation matrix $\Gamma$ serves as a dimensionality reduction operator. This enables efficient handling of high-dimensional datasets with a large number of predictive covariates while maintaining the sparsity of the factor model. This feature is particularly valuable when working with financial data (\cite{feng2020taming}) and high-dimensional macroeconomic time series data (\cite{brave2009chicago}).

Second, unlike conventional static factor models, the instrumented factor loadings in CSC-IPCA exhibit time-varying dynamics. This is particularly realistic in many economic and social science contexts. For example, consider a company that increases its investment in R\&D, transitioning from a conservative stance to a more aggressive one. This change can also impact its profitability, potentially shifting it from a robust to a weaker position. As a result, the unit effect evolves along with its investment strategy. In such cases, static factor loadings fail to capture the time-varying dynamics of the company's changing fundamentals.

Last but not least, the most valuable benefit of the CSC-IPCA method is its reduced bias when unobserved covariates are present, compared to other similar methods. Instead of including covariates linearly as regressors which is a practice often criticized for model misspecification. The CSC-IPCA method incorporates covariates into the factor loadings through a mapping matrix. This approach provides a more efficient way of extracting predictive information from predictors and reducing exposure to model misspecification. Our simulation studies demonstrate that, in the presence of unobserved covariates and when the number of unobserved covariates increases, the CSC-IPCA method is the least biased among the methods considered.

The IPCA method was developed by \cite{kelly2020instrumented}, and applied by \cite{kelly2019characteristics} for predicting stock returns in the asset pricing literature. The main difference between using the IPCA method for prediction and for causal inference lies in the assumption that the transformation matrix $\Gamma$ differs between treated and control units. In the estimation process, we first use the control units to estimate the common factors $F_t$ over the entire time period. Next, we update the transformation matrix $\Gamma_{treat}$ for the treated units using data from the pre-treatment period. The subsequent step involves normalizing the common factors and the transformation matrix based on prespecified normalization restrictions. Finally, the estimated parameters are used to impute the missing counterfactuals for the treated units after the treatment, allowing us to evaluate the average treatment effect on the treated (ATT).

Similar with other factor models, we need to decide the number of latent factors. In this paper, we provide bootstrap and leave-one-out cross-validation procedures for hyperparameter tuning to select the optimal number of latent factors, $K$. Additionally, we construct confidence intervals using the novel and increasingly popular conformal inference method developed by \cite{chernozhukov2021exact}. In our formal results, we derive the asymptotic properties based on the unbiased and efficient estimation of both $\Gamma$ and $F_t$. We show that the convergence rate of our estimand, i.e. the ATT, is the smaller of $\mathcal{O}p(\sqrt{N_{ctrl}})$ and $\mathcal{O}p(\sqrt{N_{treat}T_{pre}})$, so large $T_{pre}$ and $N_{ctrl}$ would be necessary for us to get the accurate estimation.

In the empirical application, we apply this newly developed method to assess the impact of Brexit on foreign direct investment (FDI) in the U.K. We use 9 covariates with predictive power over FDI, including GDP, GDP per capita, imports, exports, employment, and demographic data. Our empirical results indicate that Brexit had a negative effect on FDI in the U.K., and the findings are highly robust based on conformal inference. However, the prediction accuracy deteriorates after 2020, primarily due to the impact of COVID-19, which significantly altered some of the covariates.

\section{Literature Review} 
\label{sec: literature}
Causal inference in economics and other social sciences is frequently complicated by the absence of counterfactuals, which are essential for evaluating the impact of a treatment or policy intervention. \cite{imbens2015causal} state that, at some level, all methods for causal inference can be viewed as missing data imputation methods, although some are more explicit than others. For instance, under certain assumptions, the matching method (\cite{abadie2006large, abadie2011bias}) explicitly imputes the missing counterfactual for treated units with meticulously selected controls. The DID method (\cite{card1993minimum, ashenfelter1978estimating}), on the other hand, implicitly imputes the missing counterfactual by differencing the control and treated units before and after treatment. Meanwhile, the SCM method explicitly imputes the missing counterfactual with a weighted average of control units. Our method aligns with the recent trend in the causal inference literature, aiming to explicitly impute the missing counterfactuals by modeling the entire DGPs, a strategy highlighted by \cite{athey2021matrix} with their matrix completion (MC) method, and \cite{xu2017generalized} with their CSC-IFE method.

As a newly developed branch of causal inference, modeling entire DGPs offers distinct advantages. This approach helps to overcome the constraints imposed by untestable and stringent assumptions, such as unconfoundedness and common support in matching methods (\cite{rosenbaum1983central, rubin1997estimating}), as well as the parallel trends assumption in difference-in-differences (DID) models (\cite{card1993minimum}). Additionally, it addresses the limitations of the original synthetic control method (SCM) developed by \cite{abadie2006large} and \cite{abadie2010synthetic}, as well as its variants advanced by \cite{ben2021augmented} and \cite{arkhangelsky2021synthetic}, which require the outcomes of treated units to lie within or near the convex hull formed by the control units.

Factor models have long been explored in the econometrics literature related to modeling panel data, with significant contributions by \cite{bai2003computation}, \cite{pesaran2006estimation}, \cite{stock2002forecasting}, \cite{eberhardt2009cross}, among others. However, within the context of causal inference, \cite{hsiao2012panel} stands out as the first work proposing the use of these methods specifically for predicting missing counterfactuals in synthetic control settings, followed by \cite{gobillon2016regional}, \cite{xu2017generalized}, \cite{chan2016policy}, and \cite{li2018inference}. Conventional factor models with static factor loadings fail to capture time-varying factor loadings that arise due to changes in a unit's fundamentals. \cite{kelly2020instrumented} was the first to incorporate time-varying factor loadings by instrumenting them with covariates. The IPCA method has been successfully applied to stock return prediction by \cite{kelly2019characteristics}, demonstrating significant accuracy in out-of-sample predictions. Our paper is the first to apply this method for causal inference within the relevant literature.

This paper is structured as follows. Section \ref{sec: framework} introduces the framework of the CSC-IPCA method, detailing the functional form and assumptions for identification. Section \ref{sec: estimation} outlines the estimation procedures, including hyperparameter tuning and inference. Section \ref{sec: simulation} presents the results of Monte Carlo simulations, comparing different estimation methods and providing finite sample properties. Section \ref{sec: application} demonstrates the application of the CSC-IPCA method in a real-world setting, evaluating the impact of Brexit on foreign direct investment (FDI) in the U.K. Section \ref{sec: conclusion} concludes the paper with a summary of the main findings and potential future research directions. More technical details and derivations are provided in Appendix from \ref{sec: tech details} to \ref{sec: application app}.

\section{Framework} 
\label{sec: framework}

\subsection{Set up and notation} 
\label{sec: set up}
Consider $Y_{it}$ as the observed outcome for a specific unit $i \ (i = 1, \dots, N)$ at time $t \ (t = 1, \dots, T)$. The total number of observed units in the panel is $N = N_{treat} + N_{ctrl}$, where $N_{treat}$ represents the number of units in the treatment group $\mathcal{T}$ and $N_{ctrl}$ represents the number of units in the control group $\mathcal{C}$. Each unit is observed over $T = T_{pre} + T_{post}$ periods, where $T_{pre}$ and $T_{post}$ are the number of periods before and after treatment. We observe the treatment effect at $T_{pre} + 1$ right after the beginning of the treatment and continue to observe thereafter until the end of the observation periods, a scenario commonly referred to as block assignment\footnote{We can also adopt this method for the more commonly observed staggered adoption scenario. We demonstrate different treatment assignment mechanisms in Appendix \ref{app: treatment assignment}}. Following Equations \ref{eqn: fe} and \ref{eqn: instrument}, we assume that the outcome of interest $Y_{it}$ is given by a simple factor model with factor loadings instrumented by covariates. The functional form is given by:

\begin{assumption}
    Functional form:
    \label{ass: function}
\end{assumption}
    
\begin{equation}
    \begin{cases}
    Y_{it} = D_{it} \circ \delta_{it} + \Lambda_{it}F_{t}' + \mu_{it}, \\
    \Lambda_{it} = X_{it}\Gamma + H_{it}
    \end{cases}
    \label{eqn: functional form}
\end{equation}

The primary distinction of this functional form from existing fixed effects models (\cite{gobillon2016regional, chan2016policy}) is that the factor loading $\Lambda_{it}$ is instrumented by observed covariates $X_{it}$, which makes the conventionally static factor loadings exhibit time-varying features. Specifically, $F_t = [f_t^1, \ldots, f_t^K]$ is a vector of $K$ unobserved common factors, and $\Lambda_{it} = [\lambda_{it}^1, \ldots, \lambda_{it}^K]$ represents a vector of factor loadings. Meanwhile, the vector $X_{it} = [x_{it}^1, \ldots, x_{it}^L]$ comprises $L$ observed covariates. The transformation matrix $\Gamma$, which is of size $L \times K$, maps the information from observed covariates $X_{it}$ to factor loadings $\Lambda_{it}$. This integration permits $\Lambda_{it}$ to exhibit variability across time and units, thereby introducing an additional layer of heterogeneity into the model. Another key difference from the CSC-IFE approach by \cite{xu2017generalized} is that we retain only the structural component $\Lambda_{it} F_t$ between common factors and factor loadings; the linear part of covariates $X_{it}\beta$ (as specified in Equation \ref{eqn: ife}) is excluded from the functional form. The logic behind this is that we believe the unit-specific factor loadings, instrumented by covariates, have included all the predictive information from these predictive covariates.

The remainder of the model adheres to conventional standards, where $D_{it}$ denotes a binary treatment indicator, and $\delta_{it}$ represents the treatment effect, which varies across units and over time. For computational simplicity, we assume $D_{it} = 1$ for unit $i$ in the group of treated $\mathcal{T}$ and for period $t > T_{pre}$, with all other $D_{it}$ set to $0$. The model easily accommodates variations in treatment timing by removing the constraint that treatment must commence simultaneously for all treated units. The term $\mu_{it}$ signifies the idiosyncratic error associated with the outcome variable $Y_{it}$. Additionally, $H_{it} = [\eta_{it}^1, \ldots, \eta_{it}^K]$ constitutes the vector of error terms linked to $K$ unobserved factor loadings.

Following \cite{splawa1990application} potential outcome framework (also discussed by \cite{rubin1974estimating, rubin2005causal}), we observe the actual outcome for the treated and untreated units for the entire period. If we combine the two components in Equation \ref{eqn: functional form}, we get the actual outcomes for treated and controls distinguishied by different $\Gamma$ and treatment assignments, as presented in the following:

\begin{equation}
\label{eqn: potential outcome}
\begin{cases}
      Y_{it}^1 = \delta_{it} + X_{it} \Gamma_{treat} F'_t + \epsilon_{it} & if \ i \in \mathcal{T} \ \& \ t > T_{pre} \\
      Y_{it}^0 = X_{it} \Gamma_{treat} F'_t + \epsilon_{it} & if \ i \in \mathcal{T} \ \& \ t \leq T_{pre} \\
      Y_{it}^0 = X_{it} \Gamma_{ctrl} F'_t + \epsilon_{it} & if \ i \in \mathcal{C}.
\end{cases}
\end{equation}

where Equation \ref{eqn: potential outcome} represents the actual outcome for the treated and control units combined the two parts together in Equation \ref{eqn: functional form}. Our goal is to impute the missing counterfactual $\hat{Y}_{it}^0 = X_{it} \hat{\Gamma}_{treat} \hat{F}_t$ for the treated units $i \in \mathcal{T}$ when $t > T_{pre}$, where the $\hat{\Gamma}_{treat}$ and $\hat{F}_t$ are estimated parameters. We then calculate the ATT as the difference between the actual outcome and the imputed missing counterfactuals, which is defined as:

\begin{equation}
\widehat{ATT}_{t} = \frac{1}{N_{treat}}\sum_{i \in \mathcal{T}} \left( Y_{it}^1 - \hat{Y}_{it}^0 \right) = \frac{1}{N_{treat}}\sum_{i \in \mathcal{T}}\hat{\delta}_{it}.
\end{equation}

\subsection{Assumptions for identification} 
\label{sec: assumptions}
To ensure the identification of the treatment effect, we introduce a series of assumptions about the DGPs, in addition to the assumption regarding the functional form. These assumptions are crucial for achieving consistent estimation of the ATT, yet they are easy to satisfy in practice.

\begin{assumption}
    Assumption for consistency:
    \begin{enumerate}
        \item Covariate orthogonality: $\mathbb{E}\left[X'_{it} \epsilon_{it}\right] = \textbf{0}_{L\times 1}$,
        
        \item Moment condition of factors and covariates. The following moments exist: $\mathbb{E}\|F_{t}F'_{t}\|^2$, $\mathbb{E}\|X'_{it}\epsilon_{it}\|^2$, $\mathbb{E}\|X'_{it}X_{it}\|^2$, $\mathbb{E}\left[\|X'_{it}X_{it}\|^2\|F_{t}F'_{t}\|^2 \right]$, 
    
        \item Almost surely, $X_{it}$ is bounded, and define $\Omega_t^{xx} := \mathbb{E}\left[ X_{it}' X_{it} \right]$, then almost surely, $\Omega_t^{xx} > \epsilon$ for some $\epsilon > 0$.
        
        \item The parameter space $\Psi$ of $\Gamma$ is compact and away from rank deficient: $\det{\Gamma' \Gamma} > \epsilon$ for some $\epsilon>0$,
        
    \end{enumerate}
    \label{app: ass consistency}
\end{assumption}

The assumptions outlined above serve as regularity conditions necessary for ensuring the consistency of the estimator. The first condition reflects the exclusion restriction commonly associated with instrumental variable regression, indicating that the instruments (in this case, covariates) are orthogonal to the error terms. This orthogonality ensures that the covariates do not correlate with any unobserved error terms influencing the outcome. Similarly, the functional form described in Equation \ref{eqn: functional form} aligns with the framework utilized in two-stage instrumental regression. The second condition imposes a limitation on the second moment of a series of random variables, ensuring that their variances remain finite and do not escalate indefinitely. The third condition mandates that the parameter space for the mapping matrix $\Gamma$ to be compact, thereby circumventing issues related to rank deficiency. This compactness is vital for the estimability of $\Gamma$ and the existence of its inverse, mirroring standard factor analysis assumptions.

To derive the asymptotic properties of the CSC-IPCA estimator, we need to introduce additional assumptions. Assumption \ref{app: ass normality} encompasses panel-wise and cross-sectional central limit theorems for various variables, which are fulfilled by diverse mixing processes. These conditions are pivotal for determining the asymptotic distribution of common factors $F_t$ and mapping matrix $\Gamma$ estimations. For an in-depth discussion, we refer to \cite{kelly2020instrumented}. The requirement of bounded dependence stipulates that both time series and cross-sectional dependencies of $X_{it}\epsilon_{it}$ are bounded, a crucial step for establishing asymptotic normality. Meanwhile, the assumption of cross-sectional homoskedasticity simplifies the expressions for the asymptotic variances of the estimated mapping matrix $\Gamma$. It should be noted that relaxing this assumption would not alter the convergence rate.

\begin{assumption}
    Assumptions for asymptotic normality:
    \begin{enumerate}
        \item $\text{As } N, T \to \infty, \: \frac{1}{\sqrt{NT}} \sum_{i,t} \text{vect}\left( X'_{i,t} \epsilon_{i,t} F'_{t} \right) \xrightarrow{d} \text{Normal} \left(0, \Omega^{x\epsilon f} \right)$,
        
        \item $\text{As } N \to \infty, \: \frac{1}{\sqrt{N}} \sum_{i} \text{vect}\left( X'_{i} \epsilon_{i} \right) \xrightarrow{d} \text{Normal} \left(0, \Omega^{x\epsilon} \right) \: \text{for} \: \forall t$,
        
        \item $\text{As } N, T \to \infty, \: \frac{1}{\sqrt{T}} \sum_{t} \text{vect}\left( F_{t}F'_{t} - \mathbb{E}[F_{t}F'_{t}] \right) \xrightarrow{d} \text{Normal} \left(0, \Omega^{f} \right)$.
        
        \item Bounded dependence: $\frac{1}{NT} \sum_{i,j,t,s}\|\tau_{ij, ts}\| < \infty$, where $\tau_{ij, ts} := \mathbb{E} \left[ X'_{it} \epsilon_{it} \epsilon'_{js} X_{js} \right]$
        
        \item Constant second moments of the covariates: $\Omega_t^{xx} = \mathbb{E}\left[ X_{t} X'_{t} \right]$ is constant across time periods.
    \end{enumerate}
        \label{app: ass normality}
\end{assumption}

\section{Estimation} 
\label{sec: estimation}
The CSC-IPCA estimator of the treatment effect for a treated unit $i$ at time $t$ is defined as the difference between the observed outcome and its imputed counterfactual: $\hat{\delta}_{it} = Y_{it}^1 - \hat{Y}_{it}^0$. To combine the functional form in Equation \ref{eqn: functional form}, we get the structural component of the DGPs for outcome $Y_{it} = (X_{it}\Gamma) F'_{t}$. The CSC-IPCA method is estimated by minimizing the sum of squared residuals of the following objective function:

\begin{equation}
\label{eqn: obj}
\underset{\Gamma, F_t}{\arg\min} \sum_{i} \sum_{t}\left( Y_{it} - (X_{it}\Gamma) F'_{t} \right)\left( Y_{it} - (X_{it}\Gamma) F'_{t} \right)'.
\end{equation}

Unlike the IFE method (\cite{bai2009panel,xu2017generalized}), our approach requires estimating only two parameters, $\Gamma$ and $F_t$, simplifying the process. Different from principal component analysis (\cite{jolliffe2002principal,stock2002forecasting}), our method involves using covariates to instrument the factor loadings. This necessitates the estimation of $\Gamma$ rather than $\Lambda_i$, so we can not directly use eigenvalue decomposition. While the objective function in Equation \ref{eqn: obj} formulates the problem as minimizing a quadratic function with a single unknown variable (e.g., $\Gamma$) while holding the other variable (e.g., $F_t$) constant. This structure enables the application of the alternating least squares (ALS) method for a numerical solution. Generally, the imputation for the missing counterfactual $\hat{Y}_{it}^0$ is executed in four steps:

\textbf{Step 1:} The initial step entails estimating the time-varying factors $\hat{F}_t$ and the mapping matrix $\hat{\Gamma}_{\text{ctrl}}$ utilizing the ALS algorithm, with the control group data exclusively for the whole period.

\begin{equation}
(\hat{\Gamma}_{ctrl}, \hat{F_t}) = \underset{\Gamma, F_t}{\arg\min} \sum_{i \in \mathcal{C}} \sum_{t \leq T}\left( Y_{it} - (X_{it}\Gamma) F'_{t} \right)\left( Y_{it} - (X_{it}\Gamma) F'_{t} \right)'.
\label{eq: optimization}
\end{equation}

\textbf{Step 2:} The subsequent step involves estimating the mapping matrix $\hat{\Gamma}_{treat}$ for treated unit $i$ at time $t$, employing the previously estimated time-varying factors $\hat{F}_t$, using treated group data exclusively for the pre-treatment period.

\begin{equation}
\hat{\Gamma}_{treat} = \underset{\Gamma}{\arg\min} \sum_{i \in \mathcal{T}} \sum_{t \leq T_{pre}} \left( Y_{it} - (X_{it} \Gamma) \hat{F}'_{t} \right) \left( Y_{it} - (X_{it} \Gamma) \hat{F}'_{t} \right)'.
\end{equation}

\textbf{Step 3:} The third step includes normalizing the estimated mapping matrix $\hat{\Gamma}_{treat}$ and factors $\hat{F_t}$ by a set of constraints:

\begin{equation}
\label{eqn: normalization}
\begin{aligned}
\Gamma_{norm} &= \hat{\Gamma}_{treat} R, \\
F_{norm, t} &= R^{-1} \hat{F}_t, \\
s.t. \Gamma_{norm}'\Gamma_{norm} &= \mathcal{I}_K, \quad F_{norm, t} F_{norm, t}'/T = \text{Diagonal}.
\end{aligned}
\end{equation}

Similar to most factor analysis methods, the estimated $\Gamma$ and $F_t$ are not deterministic. There exists an infinite number of ``rotated'' parameters $\Gamma R$ and $R^{-1}F_t$ that yield the same objective function value (i.e., $X_{it} \Gamma F_t = X_{it} \Gamma R R^{-1} F_t$). To make the estimation identifiable, it is necessary to impose some constraints on the estimated $\Gamma$ and $F_t$. Following \cite{connor1993test, stock2002forecasting, bai2002determining}, the aforementioned restrictions reduce the model's complexity and make it easier to understand and interpret the relationships between the factors $F_t$ and factor loadings $\Lambda_{it}$. Unlike the estimation methods in \cite{bai2009panel,xu2017generalized}, where normalization constraints are set before the estimation, the structural component $X_{it} \Gamma F_t$ allows us to normalize the estimated $\Gamma$ and $F_t$ after the estimation. This enables us to easily find the rotation matrix $R$ that satisfies the above constraints.

\textbf{Step 4:} The final step involves imputing the counterfactual outcome $\hat{Y}_{it}^0$ for treated unit $i$ at time $t$ by substituting the estimated mapping matrix $\hat{\Gamma}_{norm}$ and the common factors $\hat{F}_{norm, t}$ into the following equation:

\begin{equation}
\hat{Y}_{it}(0) = (X_{it} \hat{\Gamma}_{norm}) \hat{F}'_{norm, t}, \quad \forall i \in \mathcal{T} \ \& \ T_{pre} < t \leq T.
\end{equation}

The main difference between CSC-IPCA and the instrumented principal component analysis (IPCA) as proposed by \cite{kelly2020instrumented} lies in the purpose of prediction. In the IPCA method, the authors predict the next period stock returns using all covariates from the preceding period, under the assumption that the mapping matrix $\Gamma$ remains constant across all observations. In contrast, CSC-IPCA introduces a pivotal distinction: it operates under the assumption that treated and control groups are characterized by unique mapping matrices, $\Gamma_{\text{treat}}$ and $\Gamma_{\text{ctrl}}$. This assumption is vital for the unbiased estimation of the ATT, setting the CSC-IPCA method apart by directly addressing heterogeneity in treated and control groups (also result in who receives the treatment) through the specification of group-specific mapping matrices. The detailed estimation procedures are presented in the Appendix \ref{sec: appendix estimation}.

\subsection{Hyper parameter tuning} 
\label{sec: hyperparameter}
Similar to CSC-IFE methods, researchers often encounter the challenge of selecting the appropriate number of latent factors, $K$, without prior knowledge of the true data generating process. To facilitate this selection, we introduce data-driven approaches for determining the hyperparameter $K$. Utilizing both control and treated units as training and validation data, respectively, offers a practical solution. To enhance the robustness of this process, we propose two validation methods for hyperparameter tuning. Algorithm \ref{algorithm: 1} describes a bootstrap method to ascertain $K$. This approach involves repeatedly sampling $N_{ctrl}$ control units for training data and $N_{treat}$ treated units for validation data, both with replacement. The optimal $K$ is then determined by minimizing the average sum of squared errors across these validations. We also propose a leave-one-out cross-validation method, as detailed in Appendix \ref{sec: appendix hyperparameter}.

\begin{algorithm}[!ht]
    \SetAlgoLined
    \KwData{$Y, X$}
    \KwResult{Optimal hyperparameter $k$}
    Determine the maximum possible hyperparameter $K$ and the number of repetitions $N$\;

    Initialize an array $MSE$ to store the average of sum squared error for each $k$\;
    \For{$k \leftarrow 1$ \KwTo $K$}{
        Initialize sum of squared errors: $SSE_k \leftarrow 0$\;
        \For{$n \leftarrow 1$ \KwTo $N$}{
            Construct a bootstrap training dataset $(Y^{b}_{ctrl}, X^{b}_{ctrl})$ by sampling $N_{ctrl}$ control observations with replacement\;

            Construct a bootstrap validation dataset $(Y^{b}_{treat}, X^{b}_{treat})$ by sampling $N_{treat}$ treated observations with replacement\;

            Estimate parameters $\Gamma$ and $F_t$ using the training data via the ALS method\;

            Use the estimated $\hat{\Gamma}$ and $\hat{F}_t$ to predict $\hat{Y}^{b}_{treat}$ with the validation data\;

            Compute the sum of squared error for the validation data: $SE_n \leftarrow \sum\left(Y^{b}_{treat} - \hat{Y}^{b}_{treat}\right)^2$\;

            Accumulate the sum of squared errors: $SSE_k \leftarrow SSE_k + SE_n$\;
        }
        Calculate the average sum squared error for $k$: $MSE[k] \leftarrow \frac{SSE_k}{N}$\;
    }
    Select $k$ corresponding to the minimum value in $MSE$\;
    \caption{Bootstrap Hyperparameter Tuning}
    \label{algorithm: 1}
\end{algorithm}

\subsection{Inference}
\label{sec: inference}
In the context of causal inference, the pioneering application of factor models is attributed to \cite{hsiao2012panel}. Nonetheless, a formal framework for inference using this method was not established until the works of \cite{chan2016policy} and \cite{li2018inference}. These methods of inference rely on a large number of control units and pre-treatment periods to develop asymptotic properties. In recent years, conformal inference developed by \cite{chernozhukov2021exact} has gained popularity in the causal inference literature, as seen in \cite{ben2021augmented}, \cite{roth2023s}, and \cite{imbens2024causal}. Conformal inference is a nonparametric method that provides exact and robust inference without requiring the specification of the model. Our causal inference framework, designed for predicting missing counterfactuals, allows us to construct inference procedures based on conformal prediction, as introduced by \cite{shafer2008tutorial}, to ensure robustness against misspecification. The causal effect is identified as the difference between the observed outcomes and these estimated counterfactuals, expressed mathematically as:

\begin{equation*}
\theta_{it} = Y^1_{it} - \hat{Y}^0_{it}, \quad \forall i \in N_{treat} \ \& \ T_{pre} < t \leq T,
\end{equation*}

\noindent where $\theta_{it}$ denotes the treatment effect for unit $i$ at time $t$, and $\hat{Y}^0_{it}$ represents the imputed counterfactual outcome. 

To conduct conformal inference, we first postulate a sharp null hypothesis, $H_0: \theta_{it} = \theta_{it}^0$. Under this null hypothesis, we adjust the outcome for treated units post-treatment as $\tilde{Y}_{it} = Y_{it} - \theta_{it}$. We then replace the original dataset with this adjusted part, $\tilde{Y}_{it}$. 

Secondly, we follow the estimation procedures described in Section \ref{sec: estimation} to estimate the time-varying factor $F_t$ using only control data, as before, and update $\Gamma$ for the newly adjusted treated units using the entire set of treated units\footnote{As a clarification, in the estimation section, we update $\Gamma$ using only the treated units before treatment. However, for inference, we use the entire set of treated units to update $\Gamma$.}. The concept revolves around updating $\Gamma$ using all the treated units, under the assumed null hypothesis, to minimize the occurrence of large residuals after intervention.

Thirdly, we estimate the treatment effect and compute the residuals for the treated units in the post-treatment periods. The test statistic showing how large the residual is under the null:

\begin{equation}
S(\hat{\mu}) = \left(\frac{1}{\sqrt{T_{post}}}\sum_{t > T_{pre}} |\hat{\mu}|^q \right)
\end{equation}

Where $\hat{\mu}$ represents the residual for the treated units in the post-treatment periods, we employ $q=1$ for the permanent intervention effect as designed in our study. A high value of the test statistic indicates a poor post-treatment fit, suggesting that the treatment effect postulated by the null is unlikely to be observed, hence leading to the null's rejection. 

Finally, we block permute the residuals and calculate the test statistic in each permutation. The P-value is defined as:
\begin{equation}
\hat{p} = 1 - \hat{F}(S(\hat{u})), \text{ where } \hat{F}(x) = \frac{1}{|\Pi|} \sum_{\pi \in \Pi} 1\{S(\hat{u}_\pi) < x\}.
\end{equation}

\noindent where $\Pi$ represents the set of all block permutations, the test statistic for each permutation is denoted by $S(\hat{\mu}_{\pi})$, with $x$ being the test statistic calculated from the unpermuted residuals. By employing different sets of nulls, we can compute a confidence interval at a specified confidence level.

\section{Monte Carlo Simulation} 
\label{sec: simulation}
In this section, we employ Monte Carlo simulations to assess the performance of the CSC-IPCA estimator in finite sample settings. We juxtapose the CSC-IPCA estimator against the CSC-IFE and the original SCM estimators. Our comparative analysis focuses on key metrics including bias, mean squared errors, and converage properties. 

We initiate our analysis with a DGP that incorporates $L=10$ covariates and $K=3$ common factors, along with unit and time-fixed effects:

\begin{equation}
Y_{it} = D_{i} \delta'_{t} + X_{it}\beta' + (X_{it}\Gamma) F'_{t} + \alpha_i + \xi_t + \epsilon_{it}.
\label{eq: dgp}
\end{equation}

where $X_{it} = [x_{it}^1, \ldots, x_{it}^{L}]$ denotes a vector of $L \times 1$ covariates, which follows a VAR(1) process. $X_{it} = \mu_i + A_i X_{i,t-1} + \nu_{it}$, where $A_i$ is a $ L \times L$ variance-covariance matrix\footnote{In our methodology, the variance-covariance matrix is not constrained to be diagonal, thus allowing covariates within each unit to be correlated, reflecting the typical scenario in most economic time series data. To emphasize the independence among different units, we generate $N$ unique variance-covariance matrices, each corresponding to a unit, ensuring cross-sectional independence and preserving time-series correlation. Moreover, we impose a condition on these matrices by requiring the eigenvalues of $A_i$ to have characteristic roots that reside inside the unit circle, thereby assuring the stationarity of the VAR(1) process.}, The drift term $\mu_i$ equals 0 for control units and 2 for treated units\footnote{This configuration underscores that the treatment assignment is not random; rather, it depends on the covariates $X_{it}$.}, and $\nu_{it}$ is a $L \times 1$ vector of i.i.d. standard normal errors. While $F_t = [f_t^1, \ldots, f_t^K]$ denotes the vector of common factors, adhering to a similar VAR(1) process, the variable $\epsilon_{it}$ represents the idiosyncratic error term. Unit and time fixed effects, $\alpha_i$ and $\xi_{t}$ respectively, are uniformly drawn from the interval $(0,1)$. The coefficient vector $\beta = [\beta^1, \ldots, \beta^{L}]$ associated with the covariates is drawn uniformly from $(0,1)$, and $\Gamma$, the $L \times K$ mapping matrix, is drawn uniformly from $(-0.1, 0.1)$. The treatment indicator $D_{it}$ is binary, defined as $D_{it} = 1$ for treated units during post-treatment periods, and $D_{it} = 0$ otherwise. The heterogeneous treatment effect is modeled as $\delta_{it} = \bar{\delta}_{it} + e_{it}$, where $e_{it}$ is i.i.d as standard normal, and $\bar{\delta_t} = [0, \cdots, 0, 1,2,\ldots, T_{post}]$ represents a time-varying treatment effect\footnote{Here we simplify the treatment effect to be constant across units, however the heterogeneous treatment effect across units can also be easily employed.}. Only the outcome $Y_{it}$, the covariates $X_{it}$, and the treatment indicator $D_{it}$ are observed, while all other variables remain unobserved.

\begin{figure}[!ht]
    \centering
    \caption{\textbf{CSC-IPCA Data Generating Process}}
    \includegraphics{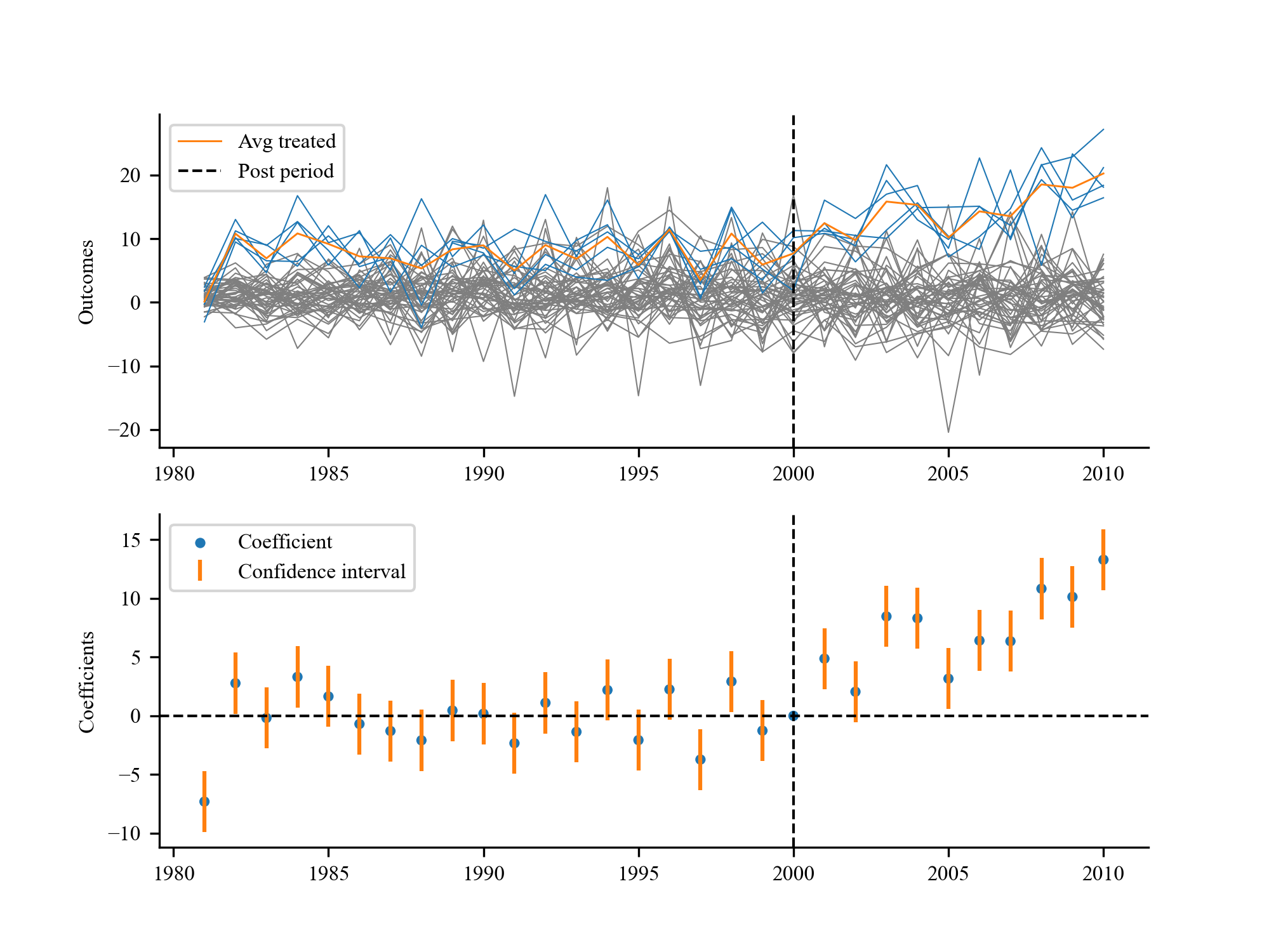}
    \label{fig: sim}
    \caption*{\footnotesize{In this graphic, the upper panel plots simulated data following the above DGPs. The light blue lines represent treated units and the light gray lines represent controls. Key parameters are $N_{treat} = 5, N_{ctrl} = 45, T_0=20, T_1=10, L=10$. The lower panel plots a simple event study.}}
    \end{figure}

Figure \ref{fig: sim} represents the simulated data following our data generating process. Observations from the upper panel indicate that the parallel trend assumption is not met. To verify this, we plot a simple event study, which clearly revealing a failure in the parallel trend assumption. Furthermore, outcomes for treated units are marginally higher than for control units. In such cases, the synthetic control method will be biased, as it avoids extrapolation and typically fits poorly for treated units.

\subsection{A simulated example}

\begin{figure}[!ht]
    \centering
    \caption{\textbf{CSC-IPCA Estimated ATT for Simulated Sample}}
    \includegraphics{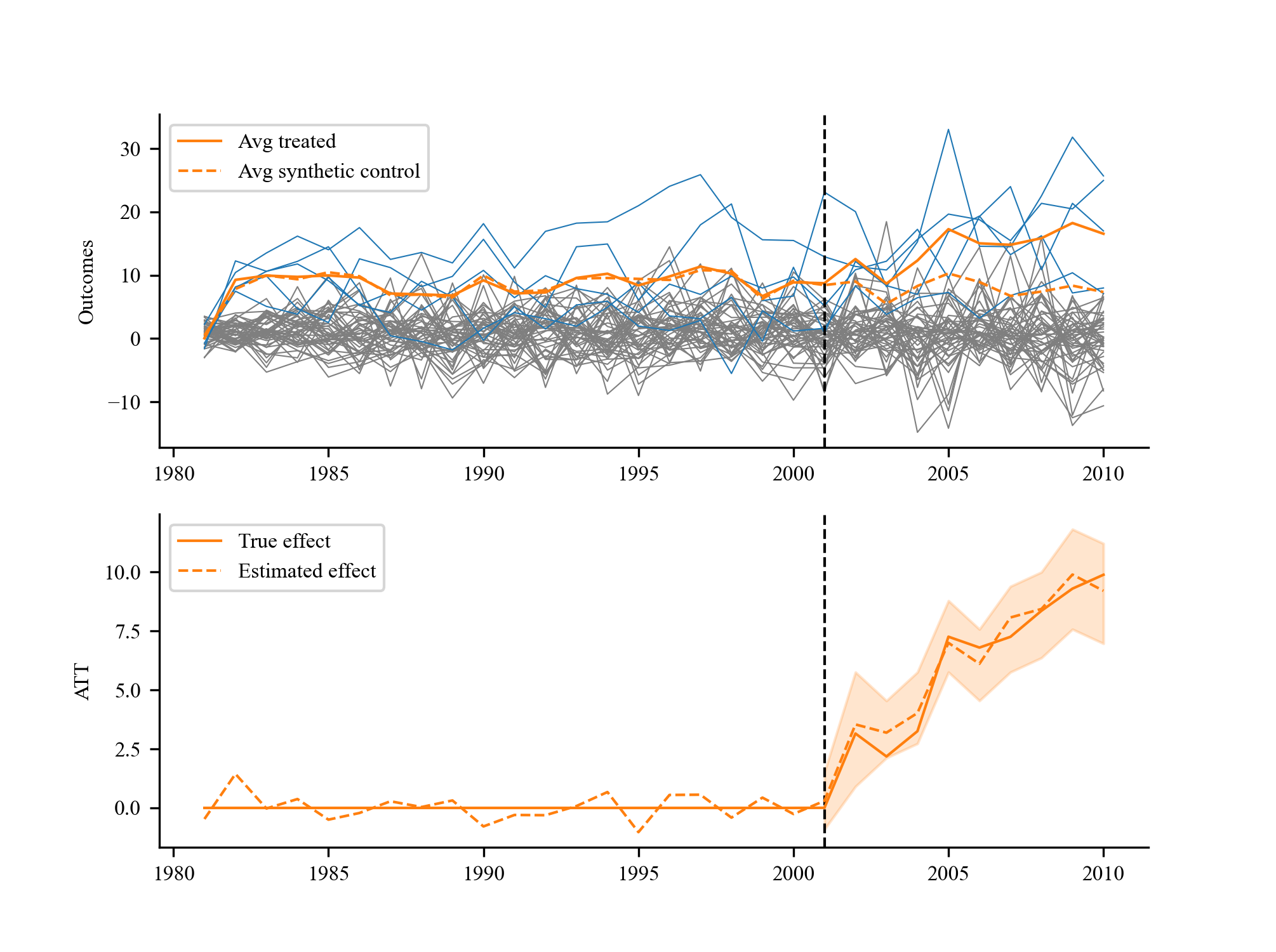}
    \label{fig: est}
    \caption*{\footnotesize{This graphic plots the CSC-IPCA method estimated ATT for simulated data $N_{treat} = 5, N_{ctrl} = 45, T_0=20, T_1=10, L=10$.}}
\end{figure}

Following this DGPs, Figure \ref{fig: est} illustrates both the raw data and the imputed counterfactual outcomes as estimated by the CSC-IPCA method. In the upper panel, control units are represented in gray and treated units in light blue, with the average outcome for treated units highlighted in orange. The imputed synthetic average for treated outcomes is also shown, delineated by an orange dashed line. The CSC-IPCA method is capable of capturing the trajectory of the average outcome for treated units before treatment, and we observe the divergence after the treatment. The lower panel of Figure \ref{fig: est} shows the estimated ATT (dashed line) alongside the true ATT (solid line) and the 95\% confidence interval based on conformal inference. The CSC-IPCA method is able to capture the true ATT, as evidenced by the close alignment between the dashed and solid lines. The confidence interval constructed through conformal inference is also accurate, as it encompasses the true ATT.

\subsection{Bias comparison}
\label{sec: bias comparing}
\begin{figure}[!ht]
\centering
\caption{\textbf{Bias Comparing with Other Methods}}
\includegraphics{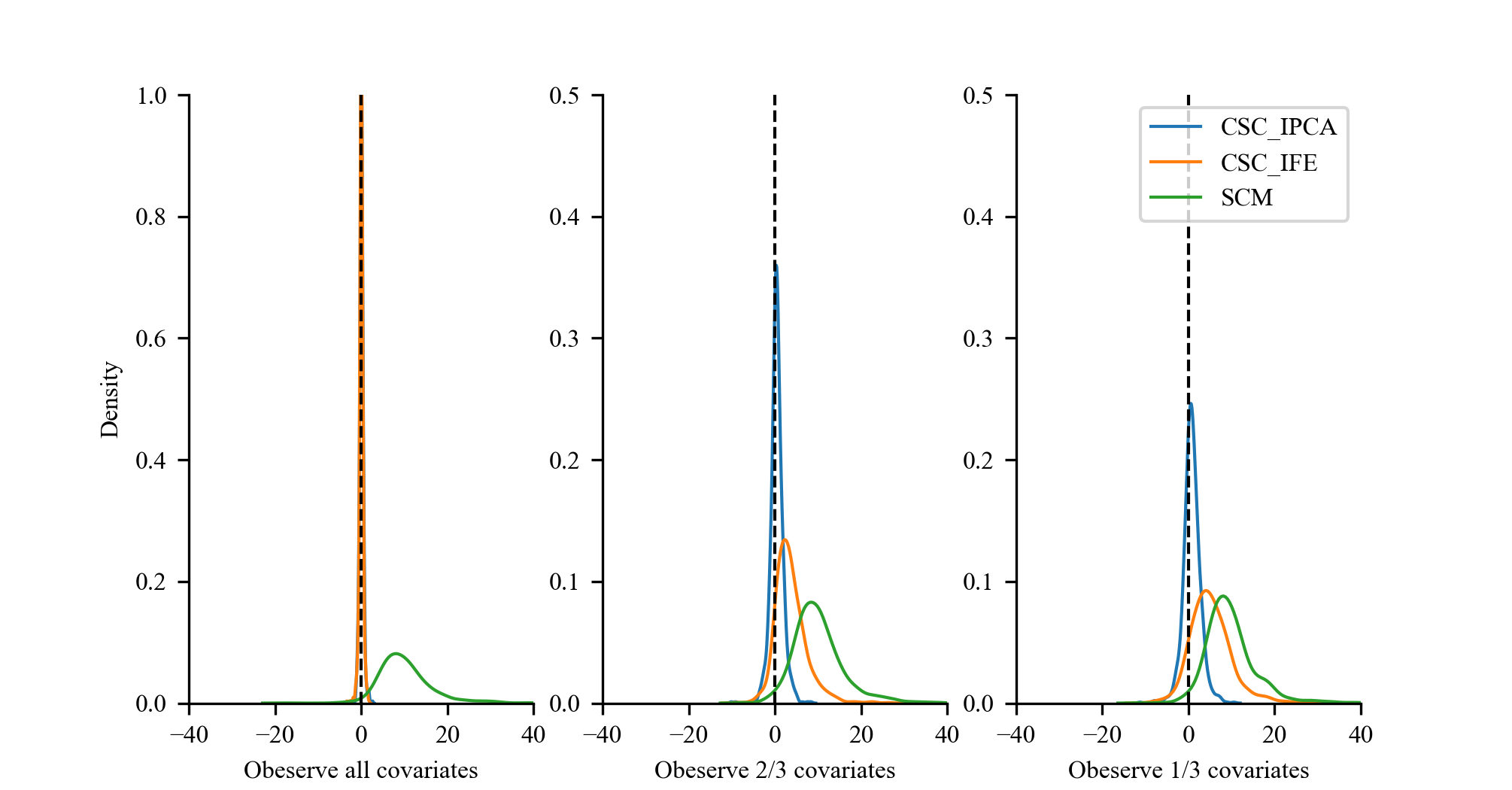}
\label{fig: bias}
\caption*{\footnotesize{This graphic plots the bias of the CSC-IPCA, CSC-IFE, and SCM estimators for simulated data $N_{treat} = 5, N_{ctrl} = 45, T_0=20, T_1=10, L=10$.}}
\end{figure}

Based on the same data generating process and parameters, we compare the CSC-IPCA, CSC-IFE, and SCM estimators with 1000 simulations. Figure \ref{fig: bias} illustrates the bias among these different estimation methods. In panel 1, when all covariates are observed, both CSC-IPCA and CSC-IFE demonstrate unbiasedness and effectively estimate the true ATT. However, due to the outcomes of treated units falling outside the convex hull of control units, the SCM exhibits an upward bias, as \cite{abadie2010synthetic} suggests that in such scenarios, SCM should be avoided. It is often the case in financial and macroeconomic studies that only a subset of covariates is observed, rather than all of them. In panels 2 and 3, we observe only 2/3 and 1/3 of the covariates, respectively. As the number of unobserved covariates increases, both CSC-IPCA and CSC-IFE lose efficiency, but the CSC-IPCA estimator remains less biased than the CSC-IFE estimator. We compare the bias of different estimators with different DGPs in Appendix \ref{app: bias 2}, and the results are consistent with the above findings.

\subsection{Finite sample properties}
\label{sec: finite sample}
We present the Monte Carlo simulation results in Table \ref{tab: finite sample} to investigate the finite sample properties of the CSC-IPCA estimator. The numbers of treated units and post-treatment periods are fixed at $N_{treat} = 5$ and $T_{post} = 5$. We vary the number of control units $N_{ctrl}$, pre-treatment periods $T_{pre}$, and the proportion of observed covariates $\alpha$ with the total number of covariates $L = 9$ to investigate the finite sample properties. As shown in Table \ref{tab: finite sample}, the bias, RMSE, and STD are estimated based on 1000 simulations\footnote{The root mean squared error (RMSE) is defined as $RMSE = \sqrt{\frac{1}{T_{pre}} \sum_{t \in T_{pre}} \left(ATT_t - \widehat{ATT}_t \right)^2}$. The standard deviation (STD) is defined as $STD = \frac{1}{T_{pre}} \sum_{t \in T_{pre}} \left(\widehat{ATT}_t - \frac{1}{T_{pre}} \sum_{t \in T_{pre}} \widehat{ATT}_t \right)^2$}. The results indicate that the bias of the CSC-IPCA estimator decreases as the lower bound, which is the smaller one between the number of control units and pre-treatment periods, increases. As suggested by the theoretical results, the convergence rate of the CSC-IPCA estimator is the smaller one between $\mathcal{O}_p\left(\sqrt{N_{ctrl}}\right)$ and $\mathcal{O}_p\left(\sqrt{N_{treat}T_{pre}}\right)$. 

In this simulation example, since the smaller one is always $\mathcal{O}_p\left(\sqrt{N_{ctrl}}\right)$, we observe that the bias decreases significantly when the number of control units increases from 10 to 40. The number of observed covariates also plays a crucial role in bias reduction. The bias decreases the most when the proportion of observed covariates increases from $1/3$ to 1 (all covariates are observed). We observe a similar pattern in RMSE and STD. It is worth noting that if we observe all the covariates (i.e., $\alpha = 1$), the bias, RMSE, and STD of the CSC-IPCA estimator all reduce to the lowest levels, even with a small number of control units and pre-treatment periods. We present the finite sample properties of the CSC-IFE and SCM estimators in Appendix \ref{app: finite sample}. When observing all the covariates, the CSC-IFE estimator is comparable to the CSC-IPCA estimator; however, when the number of observed covariates decreases, the CSC-IFE estimator becomes more biased than the CSC-IPCA estimator. The SCM estimator is always biased due to the settings.

\begin{table}[!ht]
    \centering
    \caption{\textbf{Finite Sample Properties}}
    \label{tab: finite sample}
    \begin{tabular}{cc|ccc|ccc|ccc}
    \toprule
    \multicolumn{2}{c|}{$\alpha$} & $1/3$ & $2/3$ & 1 & $1/3$ & $2/3$ & 1 & $1/3$ & $2/3$ & 1 \\
    \hline
    $T_0$ & $N_{ctrl}$ & \multicolumn{3}{c|}{Bias} & \multicolumn{3}{c|}{RMSE}  & \multicolumn{3}{c}{STD} \\
    \hline
    10 & 10 & 2.328 & 0.703 & 0.130 & 4.770 & 3.068 & 1.642 & 4.175 & 3.032 & 1.684 \\
    10 & 20 & 1.367 & 0.312 & 0.053 & 3.484 & 2.209 & 0.914 & 3.260 & 2.219 & 1.008 \\
    10 & 40 & 1.026 & 0.196 & 0.051 & 2.776 & 1.752 & 0.714 & 2.616 & 1.781 & 0.821 \\
    \cline{1-11}
    20 & 10 & 2.957 & 1.029 & 0.217 & 4.817 & 2.696 & 1.135 & 3.814 & 2.544 & 1.179 \\
    20 & 20 & 1.435 & 0.438 & 0.055 & 3.280 & 1.754 & 0.745 & 2.982 & 1.773 & 0.860 \\
    20 & 40 & 1.093 & 0.167 & 0.042 & 2.613 & 1.348 & 0.602 & 2.430 & 1.409 & 0.757 \\
    \cline{1-11}
    40 & 10 & 2.905 & 1.232 & 0.145 & 4.911 & 3.035 & 0.969 & 3.972 & 2.797 & 1.065 \\
    40 & 20 & 1.670 & 0.399 & 0.019 & 3.592 & 1.718 & 0.724 & 3.221 & 1.737 & 0.861 \\
    40 & 40 & 0.876 & 0.295 & 0.006 & 2.675 & 1.418 & 0.574 & 2.556 & 1.441 & 0.697 \\
    \bottomrule
    \end{tabular}
    \begin{tablenotes}
        \item This table presents the finite sample properties of the CSC-IPCA method estimated ATT for simulated data. The number of treated units and post-treatment period are fixed to $N_{treat} = 5$ and $T_1=5$. We vary the number of control units $N_{ctrl}$, pre-treatment period $T_0$, and proportion of observed covariates $\alpha$ to investigate the finite sample properties, the total number of covariates is $L=9$. The bias, RMSE, and STD are estimated based on 1000 simulations.
    \end{tablenotes}
\end{table}

\section{Empirical application} 
\label{sec: application}
For the empirical application, we examine the impact of Brexit on foreign direct investment (FDI) in the United Kingdom (UK). The UK's decision to leave the European Union (EU) has significantly influenced both its own economy, as discussed by \cite{arnorsson2018causes}, and the global economy, as detailed by \cite{colantone2018global}. Historically, the UK has been a major recipient of FDI, with the EU being its largest source. The 2016 Brexit referendum introduced substantial uncertainty and volatility into the UK economy, directly affecting FDI. Although several studies, such as \cite{dhingra2016impact} and \cite{welfens2018brexit}, have explored the impact of Brexit on FDI in the UK, they have not empirically examined this impact due to constraints in data and a lack of valid econometric methodology. In this paper, we aim to estimate the causal effect of Brexit on FDI to the UK using the CSC-IPCA method proposed herein.

The data used in this study is sourced from the World Development Indicators (WDI). The dataset includes information on FDI net inflows to the UK and other OECD countries, as well as other relevant variables such as GDP, imports and exports, inflation, investment, employment, and demographic indicators, the detailed description of the dataset is presented in the Appendix \ref{app: data}. The dataset covers the period from 1995 to 2022, with the Brexit referendum taking place in 2016. We define the treatment period as the post-Brexit period, starting from 2017, and the pre-treatment period as the period from 1995 to 2016. The treated unit is the UK, while the control units are the rest of the OECD countries\footnote{We also exclude countries with extreme values for FDI to GDP ratio, which may contaminate the prediction. The excluded countries include  Austria, Belgium, Hungary, Iceland, Ireland, Netherlands, Switzerland, and  Luxembourg. For detailed discussion please see \ref{app: data}}.

One of the key challenges in estimating the impact of Brexit on FDI is the high volatility of FDI data. As shown in Figure \ref{app: fdi_oecd}, FDI net inflows to each country can fluctuate significantly from year to year, making it difficult to identify the causal effect of specific events, such as Brexit. The conventional DID method may not be suitable for this type of data, as it requires the parallel trend assumption to hold, which is unlikely in this case. The synthetic control method may also fail to provide an unbiased estimate due to the high volatility of the FDI data on the one hand. On the other hand, since the treated unit is the UK, a global financial center with significant foreign investment, the FDI to GDP ratio may fall outside the convex hull of the control units, which are predominantly conventional OECD countries with less FDI compared to an international financial center like the UK. The novel CSC-IFE method is suitable for this type of data, as it can handle high volatility, lack of common support, and non-parallel trends. However, there are still caveats with the CSC-IFE method. The first concern is whether we have the correct model specification for the DGPs, since the CSC-IFE method relies on an interactive fixed effect model. The second concern is the high dimensionality of the covariates, as many variables seem to be relevant to FDI. Third, there is a issue of omitted variable bias; given the limited research on country-level FDI, there might be important variables that are not yet measured or observed. Lastly, static fixed effects may fail to capture the evolving nature of country fundamentals influenced by changes in fiscal and monetary policy, diplomatic relations, demographics, and other factors. The CSC-IPCA method can address these issues by providing a more flexible and robust estimation of the causal effect of Brexit on FDI in the UK, as discussed in the previous sections with our simulations.

\begin{figure}[!ht]
    \centering
    \caption{\textbf{CSC-IPCA Estimated ATT for Brexit}}
    \includegraphics{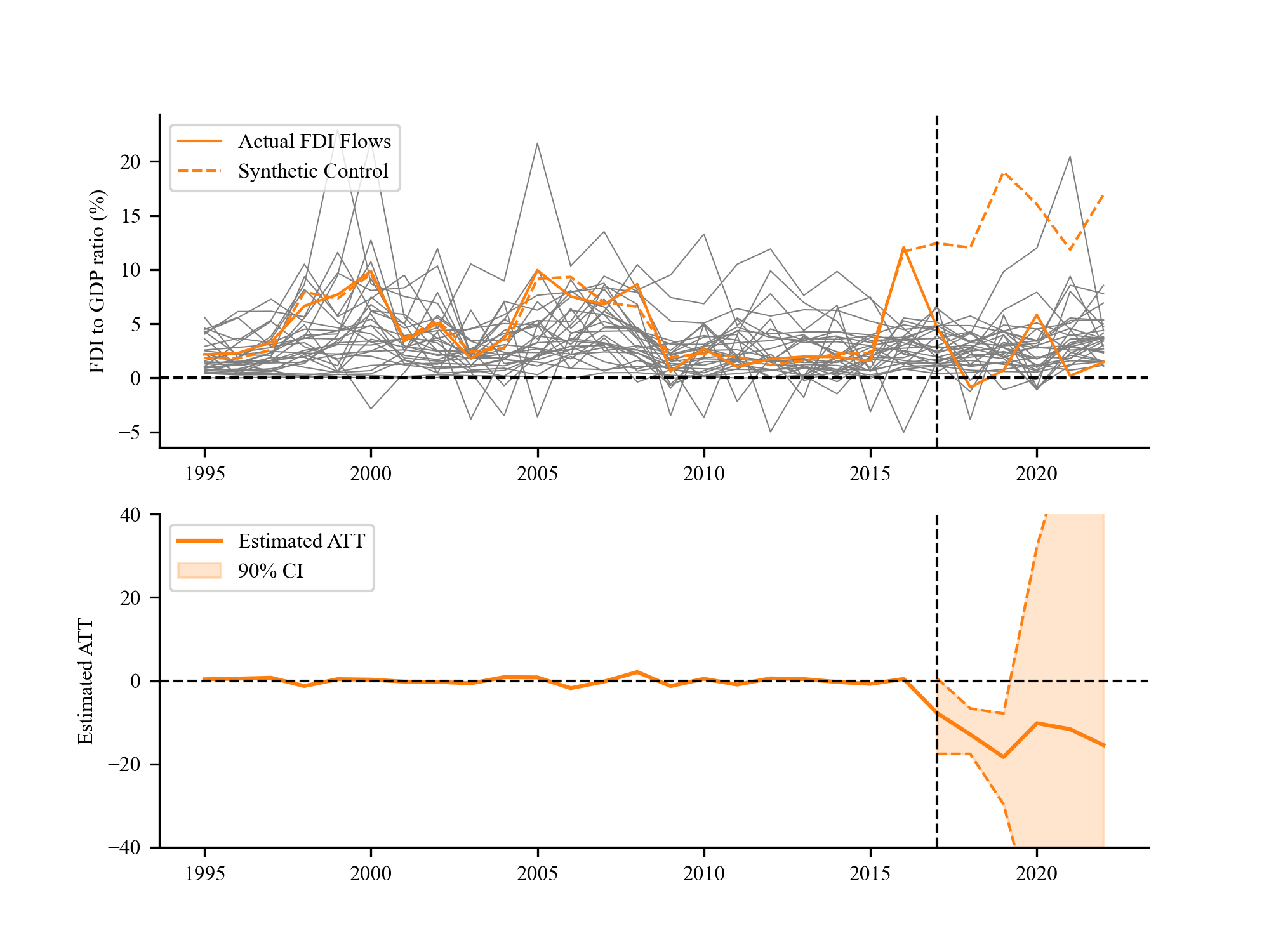}
    \label{fig: ipca_est}
    \caption*{\footnotesize{This graphic plots the CSC-IPCA method estimated ATT for the impact of Brexit on FDI in the UK.}}
    \end{figure}

For simplicity, we choose the number of latent factors $K = 2$ instead of performing hyperparameter tuning. The model converges after 52 iterations, and the estimated ATT is presented in Figure \ref{fig: ipca_est}. The upper panel of the figure shows the actual FDI net inflows to the UK and the postulated ones for the entire time period. We observe that before 2017, the postulated FDI net inflows to the UK are very close to the actual FDI net inflows, indicating that the CSC-IPCA method effectively captures the trend in FDI net inflows to the UK. However, after 2017, the postulated counterfactual FDI net inflows to the UK diverge from the actual values. 

\begin{figure}[!ht]
    \centering
    \caption{\textbf{Estimated Common Factors}}
    \includegraphics{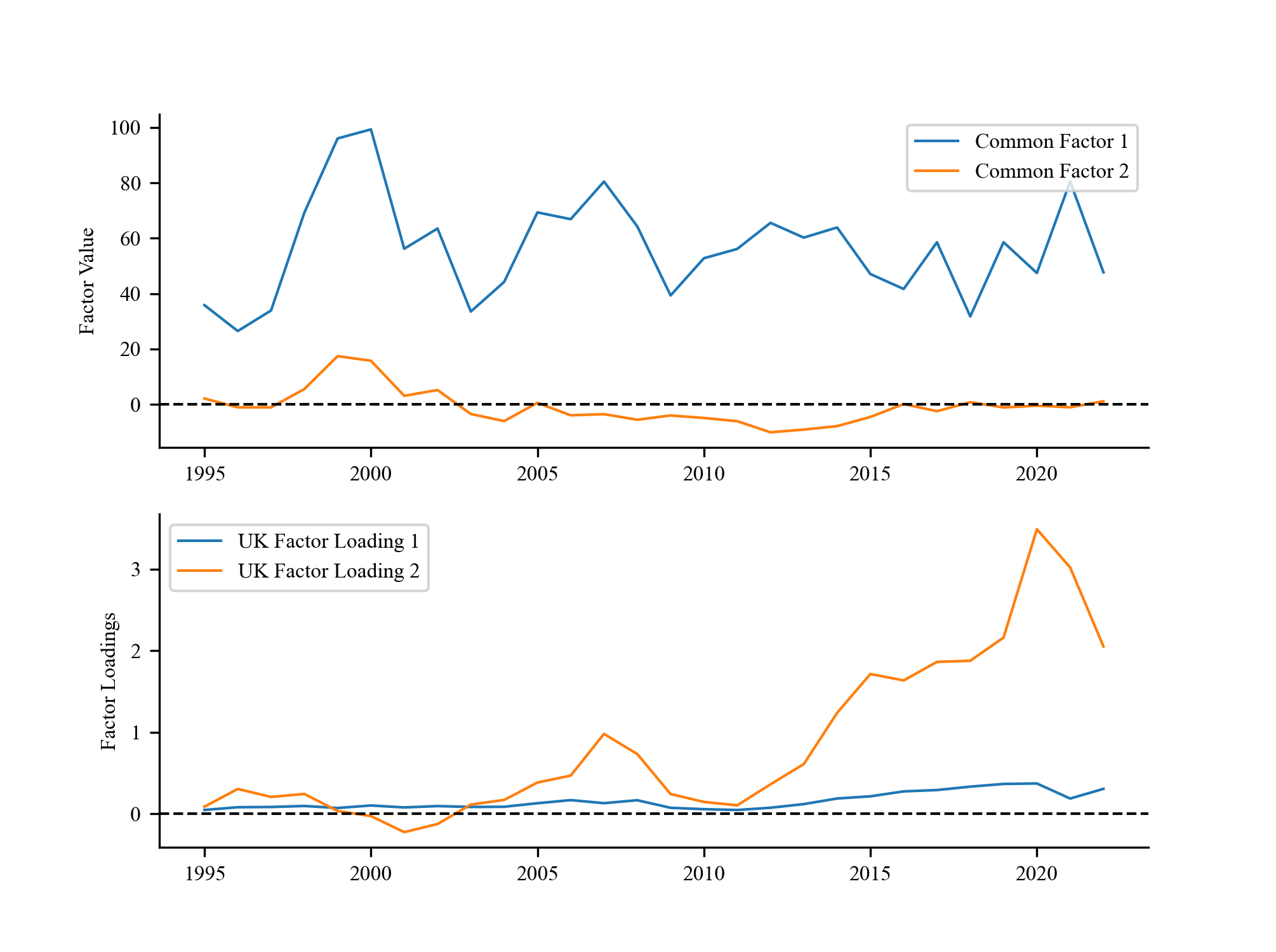}
    \label{fig: factors}
    \caption*{\footnotesize{This graphic plots the estimated common factors for FDI in OECD countries.}}
\end{figure}

The CSC-IPCA method estimates the common factors and time-varying factor loadings for the FDI data. The two estimated common factors, presented in Figure \ref{fig: factors}, are orthogonal to each other. The first common factor is more volatile, capturing the peak before the dot-com bubble burst in the late 1990s, the downturn following the bubble burst in the 2000s, and the Global Financial Crisis in 2008. The second common factor is more stable, though its interpretation is less clear. The lower panel of the figure displays the factor loadings for the UK, highlighting the time dynamics of the UK's FDI net inflows—a key difference from conventional factor models, which assume constant factor loadings over time. The first factor loading is more stable, and when it interacts with the first common factor, the interaction term primarily captures the global economic cycle's impact on FDI net inflows to the UK. In contrast, the second factor loading is more volatile, and its interaction with the second common factor captures idiosyncratic shocks to the UK's FDI net inflows.

\begin{figure}[!ht]
    \centering
    \caption{\textbf{Factor Loadings of Selected Countries}} 
    \includegraphics{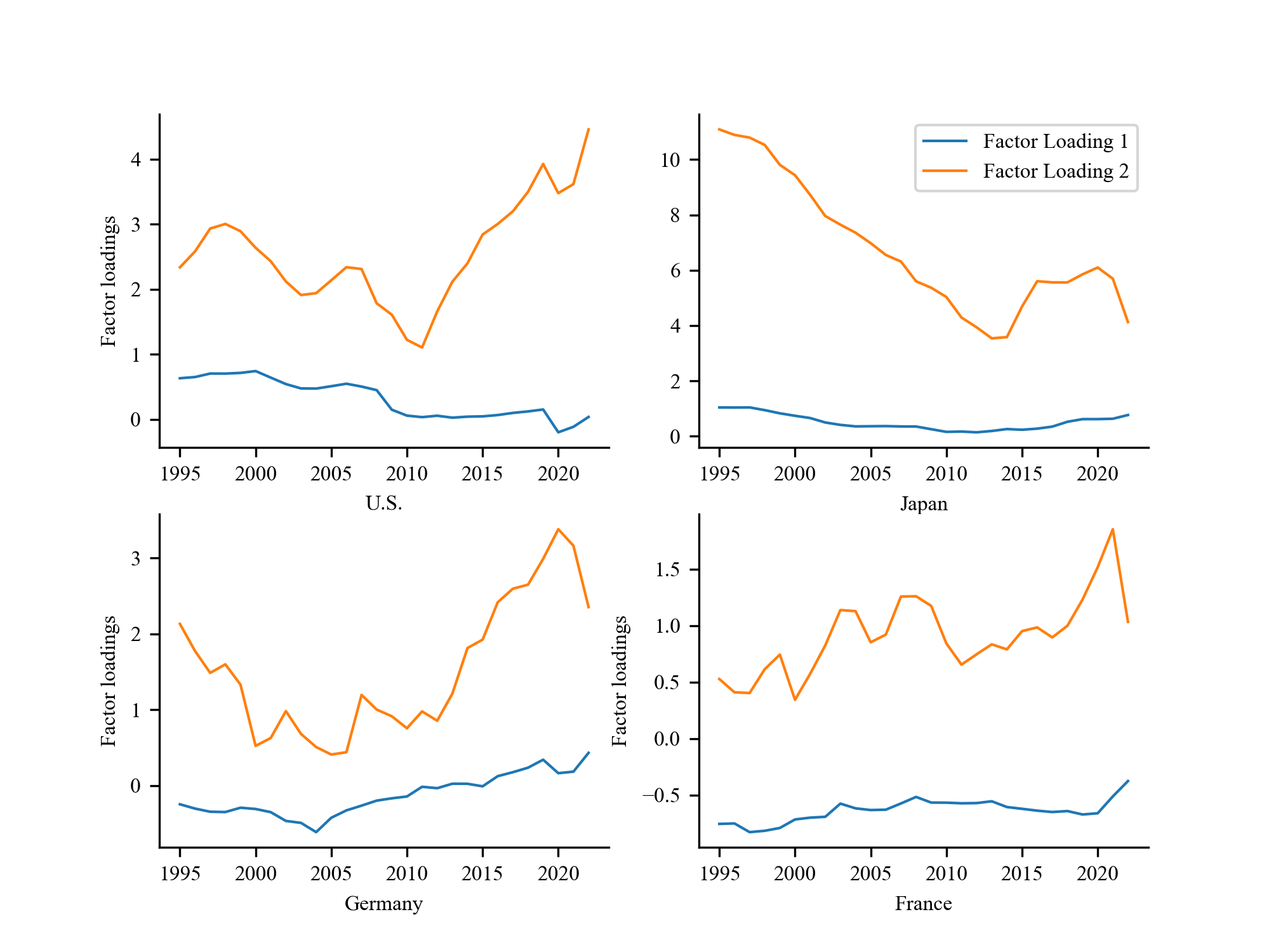}
    \label{fig: loadings}
    \caption*{\footnotesize{This graphic plots the factor loadings of selected countries.}}
\end{figure}

To further illustrate the time-varying factor loadings, we present the factor loadings for selected countries in Figure \ref{fig: loadings}. Most countries exhibit similar patterns in the relative magnitudes of the two factor loadings. We observe that the second factor loading for the US forms an approximate V-shape before and after the 2008 Global Financial Crisis, indicating a strong inflow of FDI after the crisis. The first factor loading for the US is relatively stable, similar to that of the UK. The factor loadings for Germany and France are similar, reflecting their geographic and political proximity. Japan's factor loadings display a different pattern, particularly in the second factor loading, which represents idiosyncratic shocks to Japan's FDI inflows, reflecting the prolonged struggles of the Japanese economy since the crash of the real estate bubble in the 1990s.

\section{Conclusion} 
\label{sec: conclusion}
In this paper, we introduce a new causal inference method for estimating the average treatment effect on the treated (ATT). Our approach falls within the broader family of counterfactual and synthetic control framwork and builds on the novel generalized synthetic control method, referred to here as the CSC-IFE method. This method imputes counterfactual outcomes by modeling the entire data-generating process and utilizes these imputed counterfactuals to assess the ATT.

The proposed method, CSC-IPCA, pushes the frontier of this branch of causal inference in three aspects. First, the CSC-IPCA method can easily handle high-dimensional covariates, as it employs the dimension reduction process through the mapping matrix $\Gamma$. Second, the CSC-IPCA method can capture the time-varying factor loadings, as it instruments the factor loadings with the covariates. Importantly the dynamic factor loadings have better economic interpretation and potentially can help to better estimate the common factors. Finally, by instrumenting factor loadings with covariates, the CSC-IPCA method better extracts predictive information from the data. Even in the presence of unobserved covariates, this approach reduces bias compared to other methods, enhancing the robustness of the results.

Through Monte Carlo simulations, we demonstrate that the CSC-IPCA method outperforms the CSC-IFE method, particularly in the presence of omitted variables. Our formal results and finite sample properties show that the CSC-IPCA estimator is consistent and converges rapidly to the true ATT, depending on the smaller of the number of control units or the number of pre-treatment periods. In the empirical application, we use the CSC-IPCA method to estimate the impact of Brexit on FDI in the UK. The results indicate that Brexit has a negative impact on FDI net inflows. The confidence intervals based on conformal inference are all negative before the COVID-19 pandemic, providing strong evidence to support our estimates.

One limitation of the CSC-IPCA method is the risk of overfitting when the number of latent factors is too large. While we provide two methods for selecting the number of latent factors, we suggest starting with 1 or 2 factors and increasing the number only if the pre-treatment fitting is unsatisfactory. Another concern is the potential contamination from bad controls, which we do not address in this paper. For future research, there are two key directions: first, improving techniques to manage overfitting; and second, exploring ways to address the issue of bad controls, with the goal of developing a ``kitchen sink'' model that can effectively incorporate all covariates—both good and bad—without compromising the integrity of the analysis.

\clearpage
\begingroup
\setstretch{1.0}
\bibliographystyle{plainnat}
\bibliography{citation}
\endgroup

\clearpage
\appendix
\titleformat{\section}[block]{\normalfont\Large\bfseries}{Appendix \thesection}{1em}{}
\renewcommand{\theequation}{\thesection.\arabic{equation}}
\setcounter{equation}{0}
\renewcommand{\theassumption}{\thesection.\arabic{assumption}}
\setcounter{assumption}{1}
\renewcommand{\thefigure}{\thesection.\arabic{figure}}
\setcounter{figure}{0}
\renewcommand{\thetable}{\thesection.\arabic{table}}
\setcounter{figure}{0}

\section{Technical Details} 
\label{sec: tech details}

\subsection{Treatment Assignment}
\label{app: treatment assignment}

To simplify the computation, we assume that the treatment assignment is based on a block assignment mechanism. 

\begin{figure}[!ht]
\centering
\caption{\textbf{Block Assignment}}
\includegraphics{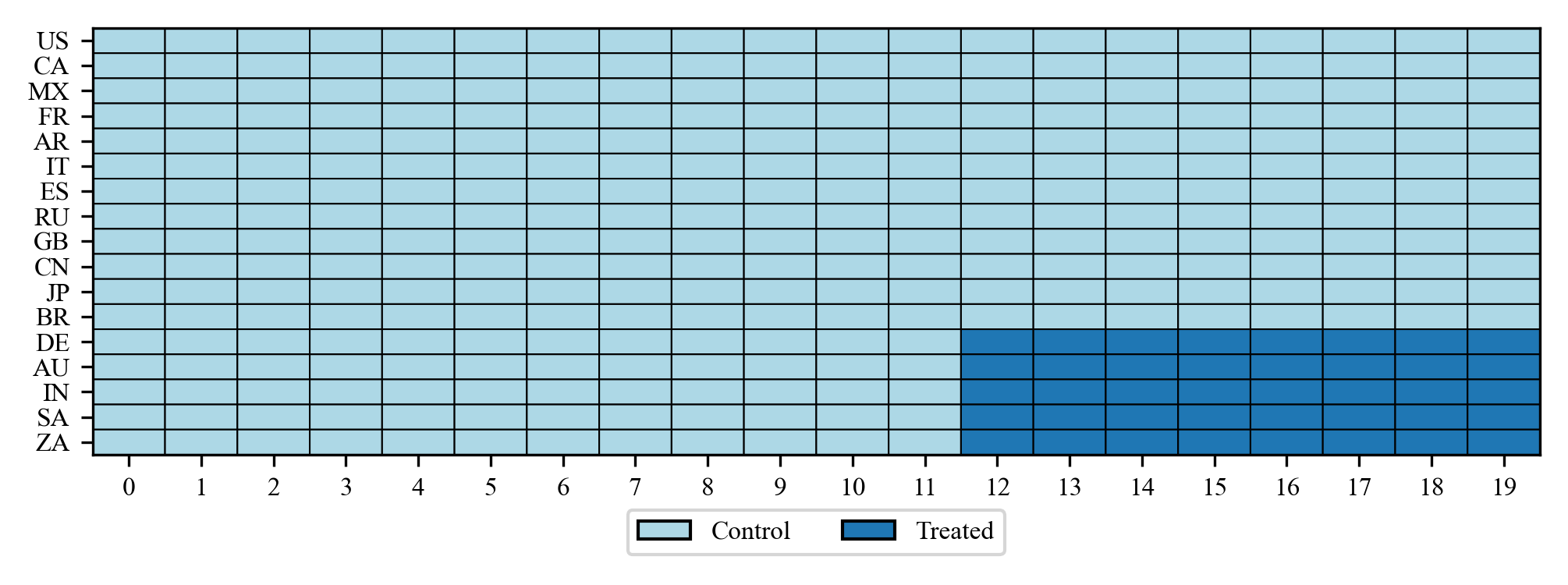}
\label{app: block assignment}
\caption*{\footnotesize{This graphic depicts the block assignment mechanism. Control units remain untreated throughout, while treated units receive the intervention simultaneously. Once the treatment is initiated, it is maintained permanently and cannot be reversed.}}
\end{figure}

The CSC-IPCA method can also be applied to estimate the staggered adoption mechanism, where treated units receive the intervention at different time points. The most complex scenario is the random assignment, where the treatment assignment is random and can be switched on or off at any time. This topic is beyond the scope of this paper, but matrix completion with nuclear norm, as detailed in \cite{athey2021matrix}, is a suitable approach for such cases.

\begin{figure}[!ht]
\centering
\caption{\textbf{Staggered Adoption}}
\includegraphics{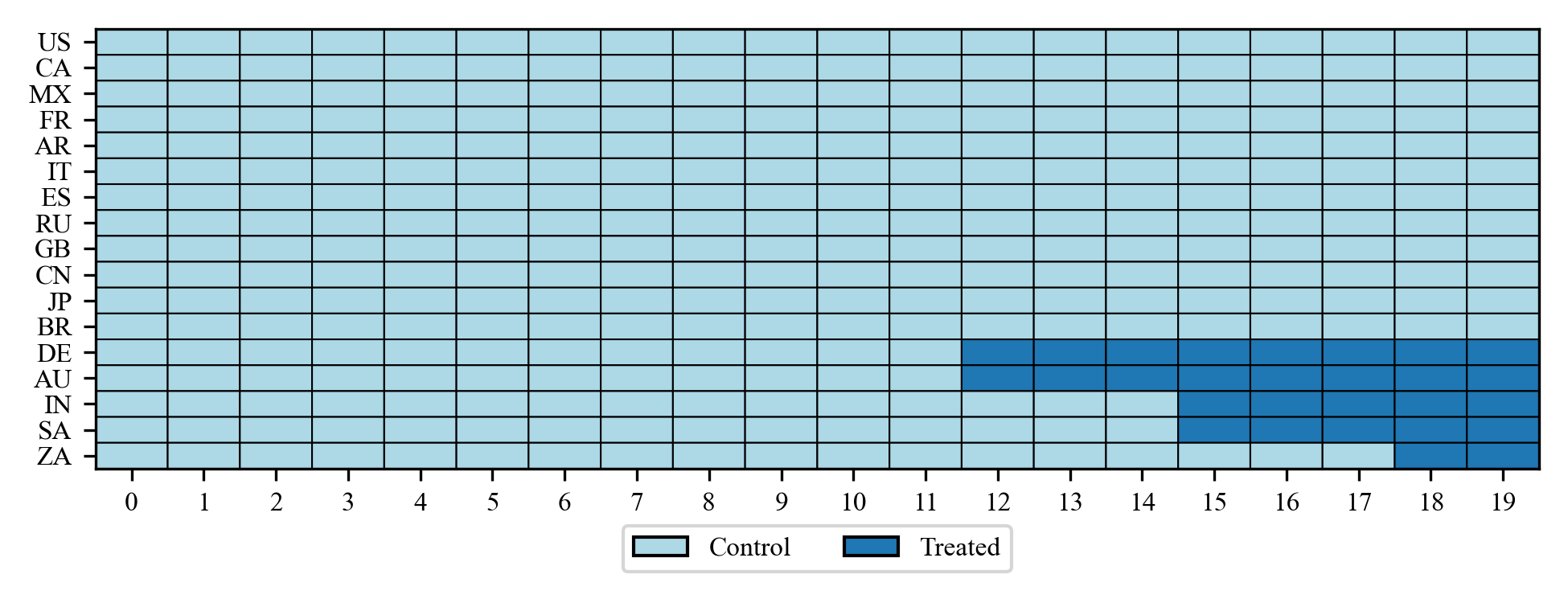}
\label{app: staggered adoption}
\caption*{\footnotesize{This graphic presents the staggered adoption mechanism. Control units remain untreated throughout, while treated units receive the intervention at different time points. Once the treatment is initiated, it is maintained permanently and cannot be reversed.}}
\end{figure}

\begin{figure}[!ht]
\centering
\caption{\textbf{Random Assignment}}
\includegraphics{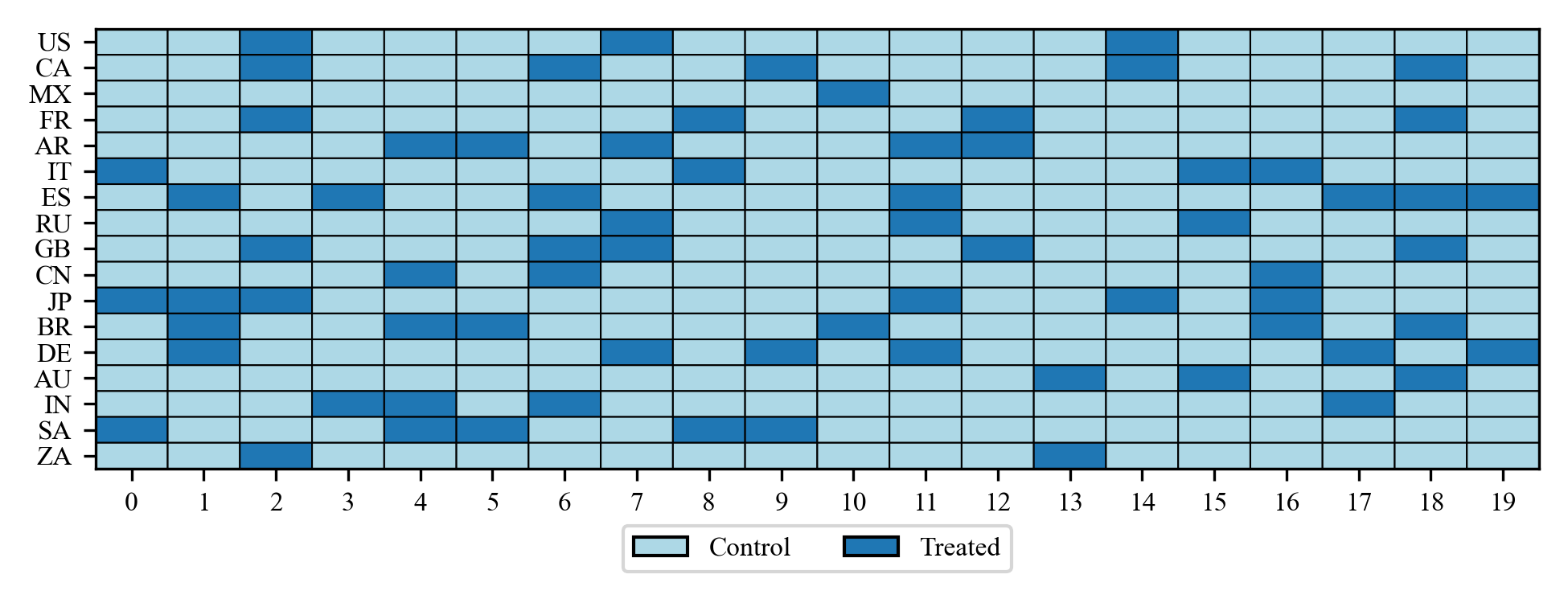}
\label{app: random assignment}
\caption*{\footnotesize{This graphic illustrates the random assignment mechanism. All units are randomly assigned to either the control or treatment group. The treatment can be switched on or off at any time.}}
\end{figure}

\subsection{Estimation of the CSC-IPCA estimator}
\label{sec: appendix estimation}
As outlined in Equation \ref{eqn: functional form}, the structural components of the data generating process are constructed by the common factors $F_t$ and dynamic factor loadings $\Lambda_{it}$, which is instrumented by the covariates $X_{it}$ through the mapping matrix $\Gamma$. The data generating process can be formulated as follows:

\begin{equation}
\label{app: eqn combined}
Y_{it} = (X_{it}\Gamma) F'_{t} + \epsilon_{it}, \quad \epsilon_{it} = \mu_{it} + H_{it} F'_t.
\end{equation}

where the error term $\epsilon_{it}$ is combined with the interaction between common factors $F_t$ and error term associated with factor loadings $H_{it}$ and the idiosyncratic error $\mu_{it}$. The objective function in Equation \ref{eqn: obj} is minimized to estimate the factor $F_t$ and mapping matrix $\Gamma$. Equation \ref{app: eqn first step} details the first step to estimate the factor $F_t$ and mapping matrix $\Gamma$ with only the control units:

\begin{equation}
\label{app: eqn first step}
(\hat{\Gamma}_{ctrl}, \hat{F}_t) = \underset{\Gamma, F_t}{\arg\min} \sum_{i \in \mathcal{T}} \sum_{t \leq T} \left( Y_{it} - (X_{it}\Gamma) F'_{t} \right)' \left( Y_{it} - (X_{it}\Gamma) F'_{t} \right).
\end{equation}

The alternating least squares (ALS) method is employed for the numerical solution of this optimization problem. Unlike PCA, the IPCA optimization challenge cannot be resolved through eigen decomposition. The optimization, as defined in the equation above, is quadratic with respect to either $\Gamma$ or $F_t$, when the other is held constant. This characteristic permits the analytical optimization of $\Gamma$ and $F_t$ sequentially. With a fixed $\Gamma$, the solutions for $F_t$ are t-separable and can be obtained via cross-sectional OLS for each $t$:

\begin{equation}
\label{app: eqn update f}
\hat{F}_t(\Gamma) = (\Gamma' X'_t X_t \Gamma)^{-1} \Gamma' X'_t Y_t.
\end{equation}

Conversely, with known $F_{t}$, the optimal $\Gamma$ (vectorized as $\bm{\gamma} = vect(\Gamma)$) is derived through pooled OLS of $Y_{it}$ against $LK$ regressors, $X_{it} \otimes F_t$:

\begin{equation}
\label{app: eqn update gamma}
\hat{\bm{\gamma}} = \left( \sum_{i,t} (X_{i,t}' \otimes F_t) (X_{i,t} \otimes F_t') \right)^{-1} \left( \sum_{i,t} (X_{i,t}' \otimes F_t) Y_{i,t} \right).
\end{equation}

Inspired by PCA, the initial guess for $F_t$ is the first $K$ principal components of the $N \times T$ outcome matrix $Y$\footnote{Here we remove the subscripte ``$it$'' indicating that matrix $Y$ represents the panel outcome of all units across all periods.}. The ALS algorithm alternates between these two steps until convergence is achieved, typically reaching a local minimum rapidly. The convergence criterion, based on the minimization of relative change in the parameters $F_t$ and $\Gamma$ in each iteration, ensures termination when this change falls below a predefined threshold, set at $1e-6$ in our implementation.

As we have mentioned before the estimation of $F_{t}$ and $\Gamma$ is not deterministic. \cite{bai2009panel} and \cite{xu2017generalized} set the constraints on the factor loadings and factors before the estimation to ensure the identifiability of the model. However, in our case, since the structural component is identified by the product between factors and factor loadings $X_{it}\Gamma F'_{t}$, we can find any arbitrary rotation matrix $R$, such that $X_{it}\Gamma R R^{-1}F'_{t}$ yields the same structural component. For a specific constraints on the mapping matrix $\Gamma_{norm} = \Gamma_{treat}R$ and factor $F_{norm, t} = R^{-1}F_t$, such that:

\begin{equation}
\begin{aligned}
& \Gamma_{norm}'\Gamma_{norm} = \mathcal{I}_K, \\
& F_{norm, t} F_{norm, t}'/T = \text{Diagonal}.
\end{aligned}
\end{equation}

where $\mathcal{I}_K$ is a $K \times K$ identity matrix, and $T$ is the number of time periods. The rotation matrix $R$ can be easily found by the following steps: first, we use Cholesky decomposition (referred to \cite{higham2009cholesky} for a guidence) to decompose the product $\Gamma' \Gamma$ into an upper triangular matrix $R_1 = cholesky(\Gamma' \Gamma)$, then we perform singular value decomposition on $R_1F_tF_t'R_1'$ to get $R_2 = U$ where $U\Sigma V'=svd(R_1F_tF_t'R_1')$. Finally, the rotation matrix $R$ is given by:

\begin{equation}
R = R_1^{-1}R_2.
\end{equation}

\subsection{Hyperparameter tuning}
\label{sec: appendix hyperparameter}
We can also utilize leave-one-out cross-validation to select the hyperparameter $K$, as detailed in Algorithm \ref{algorithm: 2}. This method involves excluding the $t^{th}$ period data from the control group to serve as the training data, while similarly excluding the corresponding period data from the treated group to act as validation data. This process is repeated for each time period in the pretreatment phase, applying a predetermined number of factor loadings. The optimal number of factors, $K$, is identified as the one that yields the minimum average of sum squared errors across all iterations.

\begin{algorithm}[!ht]
    \SetAlgoLined
    \KwData{$Y, X$}
    \KwResult{Optimal hyperparameter $k$}
    Determine the maximum possible hyperparameter $K$\;
    Initialize an array $MSE$ to store the average of sum squared error for each $k$\;
     \For{$k=1$ to $K$}{
        Set sum of squared errors $SSE_k = 0$\;
        \For{$t \leftarrow 1$ to $T_{pre}$}{
            Remove the $t^{th}$ period observation from control data, using the rest as training data $(Y_{ctrl}^{-t}, X_{ctrl}^{-t})$\;
            
            Similarly, exclude the $t^{th}$ period observation from treated data, using the rest as validation data $(Y_{treat}^{-t}, X_{treat}^{-t})$\;
            
            Estimate parameters $\Gamma$ and $F_t$ using the training data via the ALS method\;
            
            Use the estimated $\hat{\Gamma}$ and $\hat{F}_t$ to predict $\hat{Y}^{-t}_{treat}$ with the validation data\;
            
            Calculate the sum squared error $SE_t = \sum (Y_{treat}^{-t} - \hat{Y}_{treat}^{-t})^2$\;
            
            Accumulate the sum of squared errors:  $SSE_k \leftarrow SSE_k + SE_t$\;
        }
        Calculate the average sum squared error for $k$: $MSE[k] = \frac{SSE_k}{T_{pre}}$\;
      }
      Select $k$ corresponding to the minimum value in $MSE$\;
    \caption{Leave-One-Out Cross-Validation for Hyperparameter $k$}
    \label{algorithm: 2}
\end{algorithm}

\section{Formal Result} 
\label{sec: formal result}
In this section, we derive the formal result for the CSC-IPCA estimator. We first establish the consistency and other asympototic properties of the mapping matrix $\Gamma$ and the factor $F_t$, based on which we then derive the formal result for the CSC-IPCA estimated ATT. 

\subsection{Mapping matrix estimation asympototic properties}
In this section, we delve into the asymptotic properties of the estimation error associated with the mapping matrix. \cite{kelly2020instrumented} have proven it in their paper, referring to Theorem 3. The following proposition \ref{prop: gamma}, is a special case of their result. Based on our estimation methods, we estimate the mapping matrix $\Gamma$ first by concentrating out the factor $F_t$, as shown in Equation \ref{eqn: obj}, we can formulate a target function for $\Gamma$ as follows:

\begin{equation}
\label{eqn: target}
G(\Gamma) = \frac{1}{2NT}\sum_{i,t} \left( Y_{it} - X_{it}\Gamma \hat{F}_t \right)^2.
\end{equation}
we define the score function $S(\Gamma)$ as the derivative of the target function $G(\Gamma)$ with respect to $\Gamma$: $S(\Gamma) = \frac{\partial G(\Gamma)}{\partial \Gamma}$. The Hessian matrix $H(\Gamma)$ is defined as the second derivative of the target function $G(\Gamma)$ with respect to $\Gamma$: $H(\Gamma) = \frac{\partial^2 G(\Gamma)}{\partial \Gamma \partial \Gamma'}$. 

It is crucial to highlight that our normalization criterion, delineated in Equation \ref{eqn: normalization}, mandates that the mapping matrix $\Gamma$ adheres to orthonormality and the factor $F_t F_t'/T$ is required to exhibit orthogonality. To satisfy these requirements we define the following identification function:
\begin{equation}
\label{eqn: identification}
I(\Gamma) := \begin{bmatrix}
    \text{veca}(\Gamma' \Gamma - \mathcal{I}_K) \\
    \text{vecb}\left(\frac{1}{T} \sum_{t} \hat{F}_t\hat{F}_t' - V^{FF}\right)
    \end{bmatrix}
\end{equation}
where $V^{ff} = E\left[F_t F'_t\right]$, meanwhile, $\text{veca}(\cdot)$ and $\text{vecb}(\cdot)$ vectorize the upper triangular entries of a square matrix. The difference is $\text{veca}(\cdot)$ includes the diagonal elements, while $\text{vecb}(\cdot)$ excludes them. We define the Jacobian matrix $J(\Gamma)$ as the derivative of the identification function $I(\Gamma)$ with respect to $\Gamma$: $J(\Gamma) = \frac{\partial I(\Gamma)}{\partial \Gamma}$.

\begin{proposition}
\label{prop: gamma}
Under Assumption \ref{app: ass consistency} and \ref{app: ass normality}, mapping matrix estimation error centered against the normalized true mapping matrix converges to a normal distribution at the rate of $\sqrt{NT}$: as $N, T \rightarrow \infty$ such that $T/N \rightarrow \infty$,

$$
\sqrt{NT} \left( \hat{\bm{\gamma}} - \bm{\gamma}^0 \right) \xrightarrow{d} - \left( H^{0'}H^0 + J^{0'}J^0 \right)^{-1}H^{0'}Normal(0, \mathbb{V}^{[1]})
$$
\end{proposition}
where $H^0:= \frac{\partial^2 G(\Gamma)}{\partial \bm{\gamma}\partial \bm{\gamma}'}|_{\bm{\gamma} = \bm{\gamma}^0}$ and $J^0:= \frac{\partial I(\Gamma)}{\partial \bm{\gamma}}|_{\bm{\gamma} = \bm{\gamma}^0}$, $\mathbb{V}^{[1]} = \left( Q^0 \otimes \mathcal{I}_K \right) \Omega^{x\epsilon f} \left( Q^{0'} \otimes \mathcal{I}_K \right)$, and $Q^0 := Q_t(\Gamma^0)$ given that $Q_t(\Gamma) := \mathcal{I}_L - \Omega_t^{xx} \left( \Gamma' \Omega^{xx}_t \Gamma \right)^{-1}\Gamma'$ is constant over $t$ under Assumption \ref{app: ass normality}.

\textbf{Proof}: referring to \cite{kelly2020instrumented}.

\subsection{Factor estimation asympototic properties}
\begin{proposition}
\label{prop: factor}
Under Assumption \ref{app: ass consistency} and \ref{app: ass normality}, factor estimation error centered against the normalized true factor converges to a normal distribution at the rate of $\sqrt{N}$: as $N, T \to \infty$ for $\forall t$,
$$
\sqrt{N}\left(\hat{F}_t - F^0_t\right) \xrightarrow{d} N\left(0, \mathbb{V}_t^{[2]}\right),
$$
\end{proposition}
where the variance term $\mathbb{V}_{t}^{[2]}$, which is given by $\mathbb{V}_{t}^{[2]} = \left( \Gamma^\top \Omega_{t}^{xx} \Gamma\right)^{-1} \Gamma^\top \Omega_{t}^{x\epsilon} \Gamma \left(\Gamma^\top\Omega_{t}^{xx} \Gamma\right)^{-1}$.

\textbf{Proof:} Decompose the left-hand side equation:
\begin{equation*}
\begin{aligned}
\sqrt{N}\left(\hat{F}_t - F_t\right) &= \sqrt{N}\left(\left( \hat{\Gamma}'X'_tX_t\hat{\Gamma} \right)^{-1}\hat{\Gamma}'X_t' \left(X_t\hat{\Gamma}F_t+ \tilde{\epsilon}_t \right) - F_t\right)\\
&= \sqrt{N}\left(\left( \hat{\Gamma}'X'_tX_t\hat{\Gamma} \right)^{-1}\hat{\Gamma}'X_t' \left(X_t \hat{\Gamma}F_t\right) - F_t\right) + \sqrt{N}\left( \hat{\Gamma}'X'_tX_t\hat{\Gamma} \right)^{-1}\hat{\Gamma}'X_t'\tilde{\epsilon}_t
\end{aligned}
\end{equation*}
where $\tilde{\epsilon}_t$ is the estimation error term with estimated $\hat\Gamma$ and true $F_t$. Given Proposition \ref{prop: gamma}, $\hat{\Gamma} - \hat{\Gamma}^0 = \mathcal{O}_p \left( 1/\sqrt{NT} \right)$. The first term is simply $\mathcal{O}_p\left(1/\sqrt{NT}\right)$. For the second term:

\begin{equation*}
\begin{aligned}
\sqrt{N}\left( \hat{\Gamma}'X'_tX_t\hat{\Gamma} \right)^{-1}\hat{\Gamma}'X_t'\tilde{\epsilon}_t = &\sqrt{N}\left( \Gamma'X'_tX_t\Gamma \right)^{-1}\Gamma'X_t'\epsilon_t + \mathcal{O}_p(1) \\
& \xrightarrow{d} Normal(0, \mathbb{V}_t^{[2]})
\end{aligned}
\end{equation*}

Assumption \ref{app: ass normality} delineates the properties of $\Omega_t^{xx}$ and $\Omega_t^{x\epsilon}$. Notably, despite the presence of multiple matrix multiplications, the matrices $\Omega_t^{xx}$ and $\Omega_t^{x\epsilon}$ remain invariant under these operations. Consequently, the term $\mathbb{V}_t^{[2]}$ is constant across all observational units.

\subsection{Consistency and asymptotic property of the ATT estimation}
\begin{theorem}
\label{thm: bias}
Under Assumptions \ref{ass: function}, \ref{app: ass consistency}, and \ref{app: ass normality}, the CSC-IPCA estimator $\mathbb{E}\left(\widehat{ATT}_{t} | D, X, \Gamma, F\right) \xrightarrow{P} ATT_{t}$, where $ATT_{t} = \frac{1}{N_{treat}}\sum_{i \in \mathcal{T}}\delta_{it}$ is the true treatment effect. for all $t > T_{pre}$ as both $N_{ctrl}, \ T_{pre} \to \infty$.
\end{theorem}

\textbf{Proof:} Denote $i$ as the treated unit on which the treatment effect is of interest, the bias of estimated ATT is given by:

\begin{equation*}
\begin{aligned}
\hat{\delta}_{it} - \delta_{it} &= Y_{it}^1 - \hat{Y}_{it}^0 - \delta_{it}, \\    
&= X_{it}\Gamma F'_t - X_{it}\hat{\Gamma}\hat{F}'_t + \epsilon_{it}, \\
&= X_{it}\left( \left(\mathcal{I}_L\otimes F_t \right) \bm{\gamma} - (\mathcal{I}_L\otimes \hat{F}_t ) \hat{\bm{\gamma}} \right) + \epsilon_{it}, \\
&= X_{it}\left( \left(\mathcal{I}_L\otimes F_t \right) \bm{\gamma} - \mathcal{I}_L\otimes (F_t + e_{f_t}) (\bm{\gamma}+ e_{\gamma}) \right) + \epsilon_{it}, \\
&= -X_{it}\left( (\mathcal{I}_L \otimes F_t) e_{\gamma} - (\mathcal{I}_L \otimes e_{f_t} \bm{\gamma}) - (\mathcal{I}_L \otimes e_{f_t}) e_{\gamma} \right) + \epsilon_{it}\\
&= -X_{it}e_{\Gamma} F'_t - X_{it}\Gamma e'_{f_t} - X_{it}e_{\Gamma} e'_{f_t} + \epsilon_{it}, \\
&= A_{1,it} + A_{2,it} + A_{3,it} + \epsilon_{it}.
\end{aligned}
\end{equation*}

\noindent where $e_{f_t} = F_t - \hat{F}_t$ is a vector of estimation error of the factor $F_t$, $e_{\gamma}$ is vectorized estimation error of the mapping matrix $e_{\Gamma} = \Gamma - \hat{\Gamma}$. The third step converts the vector-matrix multiplication into vector multiplications with the Kronecker product, $X_{it}\Gamma F'_t = X_{it}(\mathcal{I}_L \otimes F_t)\bm{\gamma}$, where $\mathcal{I}_L$ is an $L \times L$ identity matrix. The bias of the estimated ATT is the sum of four terms $A_{1,it}$, $A_{2,it}$, $A_{3,it}$, and $\epsilon_{it}$. By proposition \ref{prop: gamma} and \ref{prop: factor}, we have the following results:
\begin{equation*}
\begin{aligned}
    &A_{1,it} = -X_{it}E_{\Gamma} F'_t = \mathcal{O}_p\left(1/\sqrt{N_{treat}T_{pre}}\right). \\
    &A_{2,it} = -X_{it}\Gamma e'_{f_t} = \mathcal{O}_p\left(1/\sqrt{N_{ctrl}}\right). \\
    &A_{3,it} = -X_{it}E_{\Gamma} e'_{f_t} = \mathcal{O}_p\left(1/\sqrt{N_{treat}T_{pre}N_{ctrl}}\right).
\end{aligned}
\end{equation*}
Since we estimate the factor $F_t$ using only control units and update the mapping matrix $\Gamma$ with treated units in the pre-treatment period, both $F_t$ and $\Gamma$ converge over different dimensions of $T$ and $N$. Consequently, the error term $\epsilon_{it}$ is assumed to have zero means, leading to the bias of the estimated ATT also converging to zero:
\begin{equation*}
\begin{aligned}
    \hat{\delta}_{it} - \delta_{it} &= \mathcal{O}_p\left(\frac{1}{\sqrt{N_{treat}T_{pre}}}\right) + \mathcal{O}_p\left(\frac{1}{\sqrt{N_{ctrl}}}\right) + \mathcal{O}_p\left(\frac{1}{\sqrt{N_{treat}T_{pre}N_{ctrl}}}\right) + \epsilon_{it} \\
    &= \mathcal{O}_p\left(\frac{1}{\sqrt{N_{ctrl}}} \right) + \mathcal{O}_p\left(\frac{1}{\sqrt{N_{treat}T_{pre}}} \right).
\end{aligned}
\end{equation*}
Therefore, as $N_{\text{ctrl}}, \ T_{\text{pre}} \to \infty$, the estimated ATT converges to the true ATT:
\begin{equation*}
\begin{aligned}
    \mathbb{E}\left(\widehat{ATT}_{t} | D, X, \Gamma, F\right) \xrightarrow{P} ATT_{t}.
\end{aligned}
\end{equation*}
The convergence rate is the smaller one between $1/\sqrt{N_{ctrl}}$ and $1/\sqrt{N_{treat}T_{pre}}$.

\section{Simulation Study} 
\label{sec: simulation app}

\subsection{Bias comparison with different DGPs}
In this section, we provide additional simulation results to compare the bias of the CSC-IPCA, CSC-IFE, and SCM estimators with different data generating processes. We consider the data generating processes used in \cite{xu2017generalized}.

\begin{figure}[!ht]
    \centering
    \caption{\textbf{Bias Comparison with Different DGPs}}
    \includegraphics{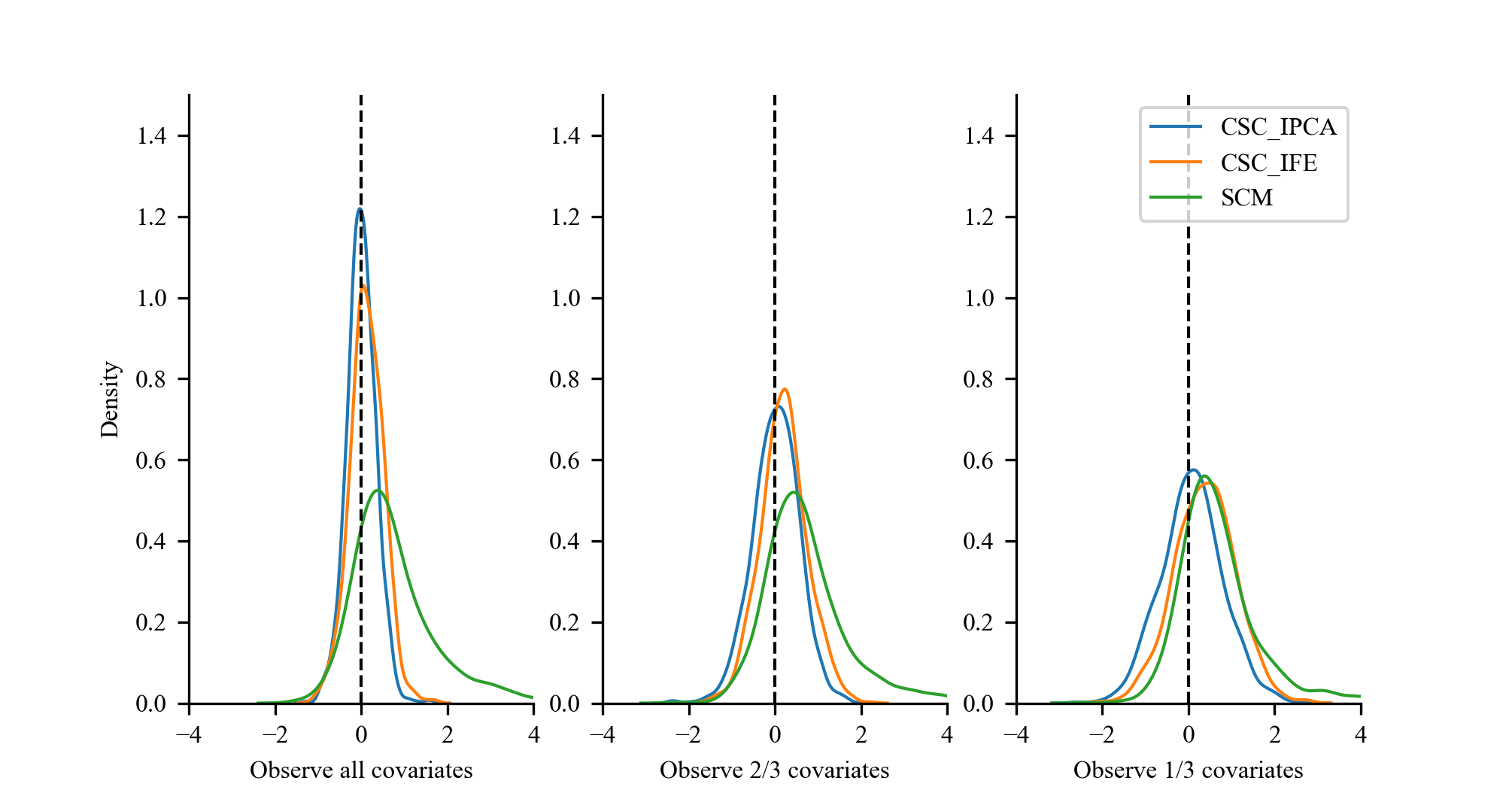}
    \label{app: bias 2}
    \caption*{\footnotesize{This figure plots the CSC-IPCA, CSC-IFE, and SCM method estimated ATT for simulated data with different data generating processes.}}
    \end{figure}

\subsection{Finite sample properties}
\label{app: finite sample}
In this section, we present additional simulation results to investigate the finite sample properties of the CSC-IFE and SCM methods for comparison. As the results show, the SCM method exhibits bias due to the specific settings of our data generating processes. While the CSC-IFE method performs comparably to the CSC-IPCA method when all predictive covariates are observed, its bias grows significantly as the number of unobserved covariates increases, leading to increasingly unreliable estimates.

\begin{table}[!ht]
    \centering
    \caption{\textbf{Finite Sample Properties of CSC-IFE}}
    \begin{tabular}{cc|ccc|ccc|ccc}
    \toprule
    \multicolumn{2}{c|}{$\alpha$} & $1/3$ & $2/3$ & 1 & $1/3$ & $2/3$ & 1 & $1/3$ & $2/3$ & 1 \\
    \hline
    $T_0$ & $N_{ctrl}$ & \multicolumn{3}{c|}{Bias} & \multicolumn{3}{c|}{RMSE}  & \multicolumn{3}{c}{STD} \\
    \hline
    10 & 10 & 6.478 & 3.184 & 0.045 & 7.996 & 4.311 & 0.747 & 4.701 & 2.939 & 0.854 \\
    10 & 20 & 6.173 & 2.885 & -0.018 & 8.070 & 4.050 & 0.599 & 5.245 & 2.900 & 0.769 \\
    10 & 40 & 4.510 & 2.516 & -0.007 & 6.785 & 3.931 & 0.527 & 5.096 & 3.044 & 0.675 \\
\cline{1-11}
    20 & 10 & 6.650 & 3.536 & -0.007 & 8.051 & 4.843 & 0.777 & 4.593 & 3.336 & 0.904 \\
    20 & 20 & 6.402 & 3.198 & -0.013 & 8.085 & 4.529 & 0.587 & 4.939 & 3.272 & 0.740 \\
    20 & 40 & 5.690 & 2.555 & 0.001 & 7.633 & 3.864 & 0.570 & 5.111 & 2.935 & 0.720 \\
\cline{1-11}
    40 & 10 & 7.353 & 3.523 & -0.008 & 11.132 & 5.258 & 0.696 & 8.364 & 3.950 & 0.846 \\
    40 & 20 & 6.904 & 3.230 & 0.036 & 9.185 & 5.084 & 0.602 & 6.053 & 3.935 & 0.747 \\
    40 & 40 & 5.978 & 2.928 & 0.003 & 9.368 & 4.850 & 0.650 & 7.227 & 3.927 & 0.782 \\
    \bottomrule
    \end{tabular}
    \begin{tablenotes}
        \item This table presents the finite sample properties of the CSC-IFE method estimated ATT for simulated data. The number of treated units and post-treatment period is fixed to $N_{treat} = 5, T_1=5$. We vary the number of control units $N_{ctrl}$, pre-treatment period $T_0$, and proportion of observed covariates $\alpha$ to investigate the finite sample properties, the total number of covariates is $L=9$. The bias, RMSE, and STD are estimated based on 1000 simulations.
    \end{tablenotes}
\end{table}

\begin{table}[!ht]
\centering
\caption{\textbf{Finite Sample Properties of SCM}}
\begin{tabular}{cc|ccc|ccc|ccc}
\toprule
\multicolumn{2}{c|}{$\alpha$} & $1/3$ & $2/3$ & 1 & $1/3$ & $2/3$ & 1 & $1/3$ & $2/3$ & 1 \\
\hline
$T_0$ & $N_{ctrl}$ & \multicolumn{3}{c|}{Bias} & \multicolumn{3}{c|}{RMSE}  & \multicolumn{3}{c}{STD} \\
\hline
10 & 10 & 10.026 & 10.188 & 9.909 & 10.997 & 11.323 & 10.964 & 4.554 & 4.960 & 4.721 \\
10 & 20 & 9.874 & 10.007 & 9.924 & 11.011 & 11.168 & 11.028 & 4.872 & 4.991 & 4.840 \\
10 & 40 & 9.720 & 10.088 & 9.521 & 10.891 & 11.388 & 10.655 & 4.936 & 5.304 & 4.808 \\
\cline{1-11}
20 & 10 & 10.596 & 10.526 & 10.674 & 12.036 & 11.841 & 12.149 & 5.714 & 5.454 & 5.816 \\
20 & 20 & 10.269 & 10.250 & 10.113 & 11.935 & 11.671 & 11.745 & 6.109 & 5.614 & 5.985 \\
20 & 40 & 9.719 & 9.654 & 10.206 & 11.309 & 11.206 & 12.353 & 5.810 & 5.699 & 6.986 \\
\cline{1-11}
40 & 10 & 10.678 & 11.069 & 11.117 & 12.794 & 12.946 & 13.705 & 7.087 & 6.731 & 8.033 \\
40 & 20 & 10.970 & 11.207 & 11.148 & 13.719 & 13.459 & 13.601 & 8.249 & 7.451 & 7.780 \\
40 & 40 & 10.742 & 10.851 & 10.221 & 13.303 & 13.786 & 12.684 & 7.853 & 8.520 & 7.539 \\
\bottomrule
\end{tabular}
\begin{tablenotes}
    \item This table presents the finite sample properties of the SCM method estimated ATT for simulated data. The number of treated units and post-treatment period is fixed to $N_{treat} = 5, T_1=5$. We vary the number of control units $N_{ctrl}$, pre-treatment period $T_0$, and proportion of observed covariates $\alpha$ to investigate the finite sample properties, the total number of covariates is $L=9$. The bias, RMSE, and STD are estimated based on 1000 simulations.
\end{tablenotes}
\end{table}

\section{Empirical Applicatoin} 
\label{sec: application app}
\subsection{Data Description}
\label{app: data}
\begin{figure}[!ht]
    \centering
    \caption{\textbf{FDI to GDP Ratio for Selected Countries}}
    \includegraphics{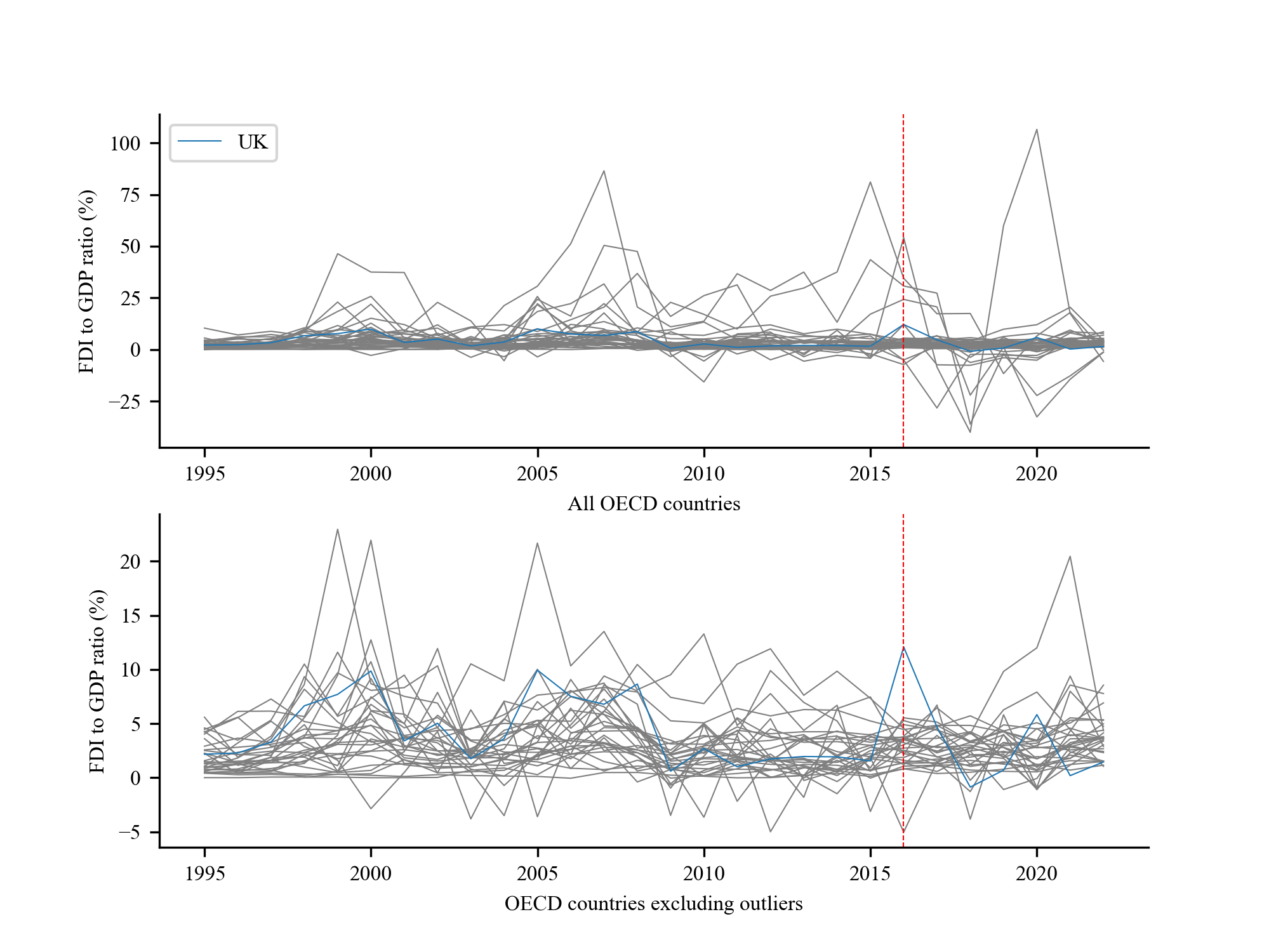}
    \label{app: fdi_oecd}
    \caption*{\footnotesize{This figure plots the FDI to GDP ratio for selected OECD countries.}}
    \end{figure}

In this section, we provide additional information on the data used in the empirical application. The main variable of interest is the FDI net inflow to GDP ratio. Based on this ratio, we exclude countries with values higher than 25\% or lower than -25\% to remove countries with extremely volatile FDI data. The excluded countries include Austria, Belgium, Hungary, Iceland, Ireland, Netherlands, Switerland and Luxembourg. As shown in Figure \ref{app: fdi_oecd}, the FDI to GDP ratio is relatively stable in most countries, with the exception of the countries mentioned above. The final sample includes 30 OECD countries from 1995 to 2022.

We provide the summary statistics of the data used in the empirical application in Table \ref{app: data summary}. The variables include log GDP, log GDP per capita, import to GDP ratio, export to GDP ratio, GDP deflator, capital formation to GDP ratio, unemployment rate, employment to population ratio, log population, and FDI to GDP ratio. 
\begin{table}[!ht]
    \centering
    \footnotesize
    \caption{\textbf{Summary Statistics of the Data}}
    \label{app: data summary}
    \begin{tabular}{lrrrrrrrr}
        \toprule
         & Count & Mean & Std & Min & 25\% & 50\% & 75\% & Max \\
        \midrule
        Log GDP & 840 & 26.65 & 1.64 & 23.05 & 25.77 & 26.46 & 27.89 & 30.69 \\
        Log GDP per Capita & 840 & 10.01 & 0.71 & 8.28 & 9.46 & 10.14 & 10.60 & 11.28 \\
        Import to GDP & 840 & 37.60 & 17.08 & 7.57 & 27.07 & 32.72 & 44.08 & 104.65 \\
        Export to GDP & 840 & 37.60 & 17.78 & 8.82 & 25.78 & 34.20 & 44.72 & 99.30 \\
        GDP Deflator & 840 & 87.53 & 34.04 & 1.29 & 73.15 & 89.95 & 100.41 & 707.41 \\
        Capital Formation to GDP & 840 & 23.60 & 4.37 & 11.89 & 20.86 & 23.08 & 25.78 & 41.56 \\
        Unemployment Rate & 840 & 8.23 & 4.24 & 2.02 & 5.11 & 7.41 & 10.28 & 27.69 \\
        Employment to Population & 840 & 55.29 & 5.80 & 37.38 & 51.56 & 56.39 & 59.49 & 68.96 \\
        Log Population & 840 & 16.64 & 1.40 & 14.09 & 15.49 & 16.35 & 17.88 & 19.62 \\
        FDI to GDP & 840 & 3.11 & 2.90 & -5.03 & 1.30 & 2.53 & 4.32 & 22.95 \\
        \bottomrule
        \end{tabular}
    \begin{tablenotes}
        \item This table presents the summary statistics of the data used in the empirical application.
    \end{tablenotes}
    \end{table}

We use different model specifications to investigate the relationship between the FDI to GDP ratio and other variables. The results are presented in Table \ref{app: reg}. The significance of the variables varies across the models, raising uncertainty about which variables are predictive and should be included in the final model. Fortunately, the CSC-IPCA method allows us to bypass the model selection process by incorporating all variables in the final model.

\begin{table}[!ht]
    \centering
    \footnotesize
    \caption{\textbf{Regression Results}}
    \label{app: reg}
    \begin{tabular}{lcccc}
        \toprule
         & Model 1 & Model 2 & Model 3 & Model 4 \\
        \midrule
        Intercept & 4.2650 & 4.2272 & -30.9338 & -51.1866 \\
         & (3.4547) & (3.3835) & (24.0543) & (35.8816) \\
        Log GDP & -0.2595*** & -0.2938*** & 0.4932 & 1.0023 \\
         & (0.0692) & (0.0674) & (0.5019) & (0.7904) \\
        Log GDP per Capita & -0.2978** & -0.3432*** & -0.4165 & -0.2885 \\
         & (0.1171) & (0.1125) & (0.7746) & (0.7574) \\
        Import to GDP & 0.0076 & -0.0088 & -0.1233*** & -0.0801* \\
         & (0.0243) & (0.0238) & (0.0409) & (0.0418) \\
        Export to GDP & 0.0179 & 0.0304 & 0.1389*** & 0.0850** \\
         & (0.0221) & (0.0214) & (0.0397) & (0.0387) \\
        GDP Deflator & -0.0056* & -0.0067* & -0.0067* & -0.0039 \\
         & (0.0030) & (0.0036) & (0.0039) & (0.0039) \\
        Capital Formation to GDP & 0.0646** & 0.0368 & 0.2544*** & 0.1152** \\
         & (0.0250) & (0.0257) & (0.0456) & (0.0491) \\
        Unemployment Rate & 0.0935** & 0.1196*** & 0.0870 & 0.0899 \\
         & (0.0370) & (0.0362) & (0.0594) & (0.0585) \\
        Employment to Population & 0.0967*** & 0.1054*** & 0.0425 & 0.0604 \\
         & (0.0259) & (0.0250) & (0.0639) & (0.0622) \\
        Log Population & 0.0383 & 0.0494 & 0.9097 & 1.2908 \\
         & (0.0838) & (0.0805) & (1.0455) & (1.1990) \\
        \midrule
        Obs & 840 & 840 & 840 & 840 \\
        R-squared & 0.1242 & 0.2245 & 0.3485 & 0.4328 \\
        R-squared Adj. & 0.1158 & 0.1907 & 0.3184 & 0.3860 \\
        Time FE & No & Yes & No & No \\
        Country FE & No & No & Yes & No \\
        TWFE & No & No & No & Yes \\
        \bottomrule
        \multicolumn{5}{l}{$^* p<0.1, \ ^{**} p<0.05, \ ^{***} p<0.01$} \\
    \end{tabular}

    \begin{tablenotes}
        \item This table presents the regression results of the FDI to GDP ratio on other variables. The standard errors are reported in parentheses.
    \end{tablenotes}
\end{table}

\end{document}